\documentclass[12pt]{iopart}
\usepackage{graphicx}
\usepackage{amsfonts}
\usepackage{amssymb}
\usepackage{cite}

\begin{document}

\title[The critical parameters enhanced by microwave field in thin
films]{The critical parameters enhanced by microwave field in
thin-film type-II superconductors}

\author{ I.V.Zolochevskii}

\address{
B.Verkin Institute for Low Temperature Physics and
Engineering, National Academy of Sciences of Ukraine, 61103
Kharkiv, Ukraine.}
\ead{zolochevskii@ilt.kharkov.ua}

\begin{abstract}
Here we give a results of experimental study of enhancement of
superconductivity by microwave irradiation in superconducting
films. An influence of the power, frequency of microwave irradiation,
as well as temperature and width of superconducting films on behavior
of experimental dependencies of enhanced critical current and the current
at which a vortex structure of the resistive state vanishes and the
phase-slip first line appears is analyzed. The experimental studies of
films with different width reveal that the effect of superconductivity
enhancement by microwave field is common and occurs in both the case of
uniform (narrow films) and non-uniform (wide films) distribution of
superconducting current over the film width. It is shown that enhancement
of superconductivity in a wide film increases not only the critical current
and the critical temperature, but also the maximum current at which the vortex
state can exist in the film. The phenomenon of superconductivity enhancement
by microwave irradiation in wide films can be described by the Eliashberg
theory, which was used to explain the same phenomenon in narrow channels.
It was experimentally found that the interval of irradiation power in which
the superconductivity enhancement was observed, became narrower with
increasing film width, what impairs the probability of its detection.

\end{abstract}

\noindent{\it Keywords\/}: enhancement of superconductivity, microwave
irradiation, critical current, wide superconducting films, phase slip line (PSL)

\submitto{Low Temp. Phys.}

\pacs{74.25.Nf, 74.40.Gh, 74.78.Db}

\maketitle

%\tableofcontents

\section{Introduction}

For a long time there was a general opinion that effect of
electromagnetic field on a superconductor should always
lead to a reduction of the energy gap $\Delta$,
the critical current $I_{c}$, the critical magnetic
field $H_{c}$ and the critical temperature $T_{c}$.
However, an increase of the critical current of a thin narrow
superconducting bridge near its critical temperature under
an influence of high-frequency electromagnetic field has
been reported in 1966 in reference \cite{WyattDmitriev}.
Later this phenomenon has been observed in almost all
types of superconducting weak links. This effect has been
explained in the Aslamazov -- Larkin theory only in
1978 \cite{AslamazovLarkin}. The phenomenon of enhancement
of superconductivity has also been found in narrow
superconducting channels (single-crystal filaments
(whiskers) and thin (the thickness
$d \ll\xi(T), \lambda_{\perp}(T)$), narrow (the
width ($d\sim\xi(T)$, $\lambda_{\perp}(T)$) films).
Here, $\xi(T)$ and $\lambda_{\perp}(T)= 2\lambda^{2}(T)/d$
are the coherence length and the
penetration depth of a weak magnetic field perpendicular
to the film, respectively, $\lambda(T)$  the London penetration depth.
In 1970 Eliashberg G.M. has proposed a microscopic theory
\cite{Eliashberg1}, which considers the effect
of electromagnetic irradiation on the energy gap $\Delta$
of a superconductor. The Eliashberg theory explained the
phenomenon of enhancement in a superconducting
channel, and this theory did not exclude a possibility
of its existence in wide films. However, decades had passed,
but this phenomenon could not been found in
wide films. After the discovery of high-temperature
cuprate superconductors (HTSC), which caused an increased
research activity, there has been a series of
works devoted to the experimental study of an
effect of microwave irradiation on
the superconducting properties of HTSC films.
Reference \cite{RudenkoKorotash} was
among the first which mentioned the finding of
the effect of superconductivity enhancement in HTSC
films. That paper shows a family of I-V characteristics
(IVC) for wide (w$\sim10~\mu$m) and long (L$\sim 15~\mu$m)
bridge of an epitaxial film of
$YBa_{2}Cu_{3}O_{7-x}$.

In the IVC it is seen that for
low ($\sim 10^{-8}$~W) power the critical current
$I_{c}$ as well as the superconducting one $I_{s}$ increase
compared to the values in the absence of microwave irradiation,
indicating the enhancement of superconductivity
by an electromagnetic field. With a further increase
in power of microwave irradiation the
critical superconducting current decreases, and harmonic and
sub-harmonic steps of the current
appear at voltages on the bridge $V_{m,n}$ related to
the frequency of an external
electromagnetic field $f$ by the Josephson relation:
$V_{mn} = (n/m)hf/2e$, where $m, n$
are integers, $h$ is Planck's constant, $e$ is the
electron charge. The authors of reference
\cite{RudenkoKorotash} suggest that the mechanism
responsible for the increase of $I_{c}$ and $I_{s}$ in
the investigated bridges of HTSC is an energy diffusion
of quasiparticles in a contact area caused by a
"jitter" of the potential well due to microwave irradiation
\cite{AslamazovLarkin}. It is this mechanism of enhancement
of superconductivity that was also found in cuprate HTSC by
other experimenters. For example, in studying the dependence
of the superconducting current
on the irradiation power in $YBa_{2}Cu_{3}O_{7-x}$ samples
\cite{Dmitriev1}.  It was found that it is the
Aslamazov -- Larkin mechanism of enhancement
\cite{AslamazovLarkin} characteristic of superconducting
weak links that is responsible for increasing the
superconducting current under microwave irradiation.
Unfortunately, preliminary studies
of the effect of superconductivity enhancement in
cuprate HTSC caused by the
Aslamazov -- Larkin mechanism \cite{AslamazovLarkin}
have not been continued.

The phenomenon of superconductivity enhancement by microwave
irradiation in quasi-one-dimensional films (narrow channels)
already belongs to classical effects in the physics of
superconductivity. The experimental manifestation of this
effect in a narrow channel is an increase of its critical
temperature $T_{c}$ and the Ginzburg -Landau critical
current $I_{c}^{GL}(T)$. When a current flowing through
a channel is greater than the current $I_{c}^{GL}(T)$,
the narrow channel comes to the resistive current state
caused solely by the appearance of phase-slip centers.
In contrast, in high-quality superconducting wide films
($w \gg \xi(T),\lambda_{\perp}(T)$) in excess of the
critical current $I_{c}(T)$ the vortex state appears,
so-called flux flow regime. A wide film is in this
regime until the transport current reaches the maximum
current $I_{m}(T)$ at which in a wide film a vortex structure
of the resistive state vanishes
\cite{AslamazovLempitsky, DmitrievZolochevskii}, and
the first phase slip line (PSL) appears. In 2001
it has been experimentally observed \cite{AgafonovDmitriev},
that in response to a microwave field not only the critical
current $I_{c}(T)$ increases but the maximum current of
existence of the vortex resistivity $I_{m}(T)$ does so.
In this connection, the problem of superconductivity
enhancement in wide films is particularly interesting,
as it requires consideration of behavior in a microwave
field of both the critical current $I_{c}(T)$ and the
maximum current of existence of the vortex resistive state $I_{m}(T)$.
\section{A microscopic theory of superconductivity of films,
enhanced by microwave irradiation}
\subsection{The effect of microwave irradiation on
the energy gap of a superconductor}
A microscopic theory of superconductivity enhancement
of films, uniform in the order parameter,
by a microwave field was proposed by Eliashberg
\cite{Eliashberg1} and developed in
references \cite{IvlevEliashberg,Eliashberg2,IvlevLisitsyn,Schmid}.
The theory applies to relatively narrow
\cite{DmitrievZolochevskiiBezuglyi} and thin
($w, d \ll \xi(T),\lambda_{\perp}(T)$) films
in which the spatial distribution of microwave
power and accordingly the enhanced gap are
uniform over the film cross section. At the same
time, the length of scattering of an electron
by impurities $l_{i}$ should be small compared
with the coherence length.

To understand properly behavior of superconductors
with an energy gap in an alternating
electromagnetic field it was necessary
to take into account both the processes of
absorption of electromagnetic energy by
quasiparticles (electrons), and the inelastic
processes of scattering of the absorbed energy.

To illustrate the physical nature of the
effect of superconductivity enhancement, we turn
to the basic equation of Bardeen -- Cooper -- Schrieffer
theory (BCS) \cite{KlapwijkBerghMooij},
which relates the energy gap $\Delta$
with the equilibrium distribution function of
electrons $n(\varepsilon)~=~(\exp^{\varepsilon/kT}+1)^{-1}$

\begin{eqnarray} \label{Eliashberg1}
\Delta=g\int_{\Delta}^{\hbar\omega_{D}}d\varepsilon
\frac{\Delta}{\sqrt{\varepsilon^{2}-\Delta^{2}}}
[1-2n(\varepsilon)]\
\end{eqnarray}

In the theory \cite{Eliashberg1}, it was shown that if
a superconductor with a uniform spatial
distribution of $\Delta$  is in an electromagnetic
field whose frequency is lower than the
frequency related with the energy gap by the
ratio $\hbar\omega=2\Delta$, and higher than the
inverse relaxation time of electrons
$\tau_{\varepsilon}$  (the relaxation time
of inelastic collisions), then the equilibrium
distribution function of electrons $n(\varepsilon)$
is shifted to higher energies, which leads to
a steady non-equilibrium state and an increase of
the superconductor's energy gap and consequently
its superconducting properties. And the
total number of excitations does not change.
This shift in the electron distribution function,
as seen in equation \eref{Eliashberg1},
results in an increase of the gap and thus enhances the
superconducting properties. A change of
$n(\varepsilon)$ is proportional to the field
intensity $E^{2}$ (for not too large $E$)
and the relaxation time of energy excitations
$\tau_{\varepsilon}$.

It should be noted that in the presence
of a microwave field the
energy gap is variable in space and time.
There is no coordinate dependence for sufficiently
thin samples. And when
$\omega\tau_{\varepsilon}\gg1$ it
turns out that temporal
oscillations of $\Delta$ can
also be ignored. It was also assumed
that the mean free path of electrons
is less than the film thickness.
Otherwise it would be necessary
to consider peculiarities of reflection from walls.

If we restrict our consideration to
not too high intensities of electromagnetic irradiation,
an equation for the time-averaged $\Delta$ is as follows:

\begin{eqnarray}\label{Eliashberg2}
\left\{\frac{T_{c}-T}{T_{c}}-
\frac{7\zeta(3)\Delta^{2}}{8({\pi k_{B} T_{c}})^{2}}-
\frac{ \pi l_{i} v_{F}e^{2}}{6T_{c}\hbar c^{2}}\left[A_{0}^{2}+
\frac{A_{\omega}^{2}}{2}-
\frac{3\Delta\hbar c^{2}}{2\pi l_{i}v_{F}e^{2}}G\right]\right\}\Delta=0
\end{eqnarray}
where $T_{c}$ is the critical temperature, $A_{0}$ is the
potential representing a static magnetic field
or a direct current, $A_{\omega}$  is the amplitude
of an electromagnetic field, $v_{F}$ is the Fermi velocity,
$l_{i}$ is the mean free path of electrons under
the scattering, $n_{1}(\varepsilon)$ is the non-equilibrium
part of $n(\varepsilon)$, $\zeta$(3)=1.202 is the
particular value of the Riemann zeta function.

\begin{eqnarray} \label{Eliashberg3}
G=-\frac{2T}{\Delta}\int_{\Delta}^{\infty}
\frac{d\varepsilon}{\sqrt{\varepsilon^{2}-\Delta^{2}}}n_{1}
(\varepsilon)
\end{eqnarray}
At low power of an external electromagnetic field
\begin{eqnarray}\label{Eliashberg4}
n_{1}(\varepsilon)=\frac{\alpha\omega}{\gamma 4T}
(\frac{\varepsilon(\varepsilon-\omega)+\Delta^{2})}
{\varepsilon\sqrt{(\varepsilon-\omega)^{2}-\Delta^{2}}}\ \theta
(\varepsilon-\Delta-\omega)-\\  \nonumber
\frac{\varepsilon(\varepsilon+\omega)+\Delta^{2}}{
\varepsilon\sqrt{(\varepsilon+\omega)^{2}-\Delta^{2}}}\ \theta
(\varepsilon-\Delta)-\\   \nonumber
\frac{2\varepsilon(\varepsilon-\omega)+\Delta^{2}}{\omega
\sqrt{(\varepsilon-\omega)^{2}-\Delta^{2}}}\
\theta(\varepsilon-\Delta)
\theta(\omega-\Delta-\varepsilon))
\end{eqnarray}
where $\alpha=(1/3)v_{F} l_{i} e^{2} A_{\omega}^{2}/\hbar c^{2}$
is proportional to the power of an
external electromagnetic field,
$\gamma = \hbar /\tau_{\varepsilon}$.
Taking into account \Eref{Eliashberg4}, \Eref{Eliashberg3}
can be written as
\begin{eqnarray}\label{Eliashberg5}
G=\frac{\omega^{2}\alpha}{2\Delta\gamma}
\int_{\Delta}^{\infty}d\varepsilon
\frac{\varepsilon(\varepsilon+\omega)
+\Delta^{2}}{\varepsilon(\varepsilon+
\omega)\sqrt{\varepsilon^{2}-\Delta^{2}
[(\varepsilon+\omega)^{2}-\Delta^{2}]}}+\\  \nonumber
\frac{1\alpha}{\Delta\gamma}
\int_{\Delta}^{\omega-\Delta}d\varepsilon
\frac{\varepsilon(\varepsilon-\omega)+
\Delta^{2}}{\sqrt{\varepsilon^{2}-\Delta^{2}
[(\varepsilon-\omega)^{2}-\Delta^{2}]}}\theta(\omega-2\Delta)
\end{eqnarray}
or
\begin{eqnarray}\label{Eliashberg6}
G=\frac{\alpha\hbar\omega}{2\gamma\Delta}
f\left(\frac{\hbar\omega}{\Delta}\right)
\end{eqnarray}
In limiting cases, the function
\begin{eqnarray}\label{Eliashberg7}
f\left(\frac{\hbar\omega}{\Delta}\right)=\frac{\hbar\omega}{\Delta}
\left[\ln\left(\frac{8\Delta}{\hbar\omega}\right)
-1\right]~  \textup{at}  ~ \frac{\hbar\omega}{\Delta}\ll1 \\
f\left(\frac{\hbar\omega}{\Delta}\right)=
\frac{\pi\Delta}{\hbar\omega}
~\textup{at}~\frac{\hbar\omega}{\Delta}\gg1   \nonumber
\end{eqnarray}
Taking into account \Eref{Eliashberg7}, \Eref{Eliashberg2}
can be rewritten as follows:
\begin{eqnarray}\label{Eliashberg8}
\frac{T_{c}-T}{T_{c}}-
\frac{7\zeta(3)\Delta^{2}}{8({\pi k_{B} T_{c}})^{2}}- \\  \nonumber
\frac{ \pi l_{i} v_{F}e^{2}A_{\omega}^{2}} {6 k_{B}T_{c}\hbar c^{2}}
\left[A_{0}^{2}+A_{\omega}^{2}\left(1-\frac{\hbar\omega}{2\pi\gamma}
f\left(\frac{\hbar\omega}{\Delta}\right)\right)\right]=0
\end{eqnarray}
In \Eref{Eliashberg8} there is no term accounting for the
interaction of an electromagnetic field
with excitations, located substantially above the
gap edge, which has the form  \cite{IvlevEliashberg}
\begin{eqnarray}
-0.11\frac{\pi}{2}\left(\frac{\hbar\omega}{k_{B}T_{c}}\right)^{2}
\frac{\alpha}{\gamma}        \nonumber
\end{eqnarray}
Now we can write a complete equation of the microscopic
theory of superconductivity, which takes into
account basic mechanisms of the interaction of
a superconductor with an external electromagnetic irradiation
\begin{eqnarray}\label{Eliashberg9}
\frac{T_{c}-T}{T_{c}}-\frac{7\zeta(3)\Delta^{2}}{8({\pi k_{B} T_{c}})^{2}}-
\frac{\pi \alpha}{2k_{B}T_{c}}\\   \nonumber
\times\left\{\frac{A_{0}^{2}}{A_{\omega}^{2}}+1+
0.11\frac{(\hbar\omega)^{2}}{\gamma k_{B}T_{c}}-
\frac{(\hbar\omega)^{2}}{2\pi
\gamma \Delta}\left[\ln\left(\frac{8\Delta}{\hbar\omega}
\right)-1\right]\right\}=0
\end{eqnarray}
In this equation, the first two terms describe a temperature
dependence of the equilibrium $(\alpha = 0)$
superconducting gap, and the third one is a contribution
of a static magnetic field or a direct current.
The fourth term of the equation describes a usual
pair-breaking effect in an external microwave field,
the fifth is a contribution of high-energy excitations,
and the sixth term is a contribution
of the interaction with an external electromagnetic field
of quasiparticles located at the Fermi surface. It is this
interaction that is responsible for the effect of
superconductivity enhancement \cite{Eliashberg1}.
Effects of heating of a superconductor by an electromagnetic
field are not considered in \Eref{Eliashberg9}.

It is important to note one more circumstance.
In \Eref{Eliashberg9} it is seen that with increasing
the radiation frequency a contribution of the latter
two terms increases, and an effect of second one
leads to an increase of $\Delta(T,\alpha)$ for a
given electromagnetic field power $\alpha$. $\Delta(T,\alpha)$
is greater than $\Delta(T,\alpha= 0)$ (superconductivity
enhancement), when
\begin{eqnarray}\label{Eliashberg10}
\left[1+0.11\frac{(\hbar\omega)^{2}}{\gamma k_{B}T_{c}}-
\frac{(\hbar\omega)^{2}}{2\pi
\gamma \Delta}(\ln\left(\frac{8\Delta}{\hbar\omega}\right)-
1)\right]\leq0
\end{eqnarray}

At not very high frequency of external irradiation the term
$0.11(\hbar\omega)^{2}/(\gamma k_{B}T_{c})$,
which describes a contribution of high-energy excitations
can be neglected, and with
$\ln(8 \Delta / \hbar\omega) > 1$  from \Eref{Eliashberg10}
we obtain an expression for the lower
frequency limit of the superconductivity enhancement

\begin{eqnarray}\label{Eliashberg11}
\omega_{L}^{2}=\frac{2\pi\gamma\Delta}{\hbar^{2}
\ln \left(8\Delta/\hbar\omega\right)}=
\frac{2\pi\Delta}{\hbar\tau_{\varepsilon}
\ln\left(8\Delta/\hbar\omega\right)}
\end{eqnarray}
\subsection{A non-equilibrium critical current of
superconducting films in a microwave field}
Theoretical studies
\cite{Eliashberg1,IvlevEliashberg,Eliashberg2,IvlevLisitsyn,Schmid},
considered superconductivity enhancement for narrow
channels in which the equilibrium energy gap and
the superconducting current density $j_{s}$ are
distributed uniformly over the sample cross section.

According to this theory, the effect of microwave
irradiation on the energy gap of a superconductor,
through which a constant transport current with
density $j_{s}$ flows, is described by
\Eref{Eliashberg9} which can be rewritten as
\begin{eqnarray} \label{Eliashberg12}
\frac{T_{c} - T}{T_{c}}-
\frac{7\zeta(3)\Delta^2}{8(\pi k_{B} T_{c})^2}-
\frac{2k_{B}T_{c}\hbar}{\pi e^2 D \Delta^4
N^2(0)}j_{s}^2 +M(\Delta)=0
\end{eqnarray}
where $N(0)$ is the density of states at the
Fermi level, $D=v_{F}l_{i}/3$ is the diffusion coefficient,
$v_{F}$ is the Fermi velocity, and $M(\Delta)$
is a non-equilibrium add-on due to the "non-equilibrium"
of the distribution function of electrons
\cite{DmitrievKhristenko,PalsRamekers,DmitrievGubankovNad}
\begin{eqnarray} \label{Eliashberg13}
M(\Delta)=-\frac{\pi \alpha}{2k_{B}T_{c}}\left[1+
0.11\frac{(\hbar \omega)^2}{\gamma k_{B}T_{c}}-\frac{(\hbar
\omega)^2}{2\pi \gamma \Delta}\left(\ln
\frac{8 \Delta}{\hbar\omega}-1\right)\right], \\  \nonumber
\hbar \omega \ll\Delta.
\end{eqnarray}
Often, an experimental study of the enhancement
effect assumes the measurement of the critical current,
rather than the energy gap. Using \Eref{Eliashberg12}
and \Eref{Eliashberg13} one can derive an
expression for the density of the superconducting current,
$j_{s}$, as a function of energy gap,
temperature and microwave power
\begin{eqnarray} \label{Eliashberg14}
j_{s}=\eta \Delta^{2}\{\frac{T_{c}-T}{T_{c}}-
\frac{7\zeta(3)\Delta^{2}}{8(\pi k_{B} T_{c})^{2}}-
\frac{\pi\alpha}{2k_{B}T_{c}}\\  \nonumber
\times \left[1+0.11\frac{(\hbar\omega)^{2}}{\gamma k_{B}T_{c}}-
\frac{(\hbar\omega)^{2}}{2\pi\gamma\Delta}\left(\ln
\frac{8\Delta}{\hbar\omega}-1\right)\right]\}^{1/2}, \\ \nonumber
\eta =eN(0)\sqrt\frac{\pi D}{2\hbar kT_{c}}
\end{eqnarray}
The extremum condition for the superconducting current,
$\partial j_{s}/\partial \Delta=0$, at
given temperature and irradiation power results in
a transcendental equation for the energy gap
$\Delta_{m}$, for which the maximum value $j_{s}$ is
reached, i.e., the critical current density
\begin{eqnarray} \label{Eliashberg15}
\frac{T_{c}-T}{T_{c}}-\frac{21\zeta(3)\Delta_{m}^{2}}{(4\pi k_{B} T_{c})^2}-
\frac{\pi \alpha}{2k_{B}T_{c}} \\  \nonumber
\times \left[1+0.11\frac{(\hbar \omega)^{2}}{\gamma k_{B}T_{c}}-
\frac{(\hbar \omega)^{2}}{4\pi
\gamma \Delta_{m}}\left(\frac{3}{2} \ln
\frac{8 \Delta_{m}}{\hbar \omega}-1\right) \right]=0.
\end{eqnarray}
Thus, substituting the solution $\Delta_{m}$ of
\Eref{Eliashberg15} into
\Eref{Eliashberg14} we find an expression for the
critical current in a microwave
field \cite{DmitrievKhristenko2}
\begin{eqnarray} \label{Eliashberg16}
I_{c}^{P}(T)=\eta d w \Delta_{ m}^2 \{\frac{T_{c}-T}{T_{c}}-
\frac{7\zeta(3)
\Delta_{m}^{2}}{8(\pi k_{B} T_{c})^{2}} -
\frac{\pi \alpha}{2 k_{B}T_{c}}   \\  \nonumber
\times \left[1+0.11\frac{(\hbar \omega)^{2}}{\gamma k_{B}T_{c}}-
\frac{(\hbar \omega)^{2}}{2 \pi \gamma \Delta_{m}}
\left( \ln \frac{8\Delta_{m}}{\hbar\omega}-1\right) \right]\}^{1/2}.
\end{eqnarray}
Without external microwave field ($\alpha=0$),
\Eref{Eliashberg16} is transformed into
an expression for the equilibrium pair-breaking current
\begin{eqnarray} \label{Eliashberg17}
I_{c}(T)= I_{c}^{GL}(T) = \eta d w \Delta_{m}^2
\left[ \frac{T_{c}-T}{T_{c}}-
\frac{7\zeta(3)\Delta_{m}^2}{8(\pi k_{B} T_{c})^2} \right]^{1/2},
\end{eqnarray}
In this case $\Delta_{m}=\sqrt{2/3} \Delta_{0}$, where
\begin{equation} \label{Eliashberg18}
\Delta_{0}(T)= \pi k_{B} T_{c} \sqrt{8(T_{c}- T)/7\zeta(3)T_{c}}=
3.062 k_{B} T_{c} \sqrt{1-T/T_{c}}
\end{equation}
is the equilibrium value of the gap at zero transport current.

Note that the use in \Eref{Eliashberg17} of the
\Eref{Eliashberg14} with the
density of states $N(0)=m^{2}v_{F}/ \pi^{2} \hbar^{3}$ ,
calculated in the free electron model,
results in a significant difference between the theoretical
and experimental values of the
equilibrium critical current, indicating a relative
roughness of such an estimation for
a metal (e.g., tin) used in preparation of samples.
At the same time, expressing the density of
states in terms of the experimentally measured quantity,
the resistance of the film on a
square, $R^{\square}=R_{4.2}w/L$, where $R_{4.2}$ is
the total film resistance
at $T = 4.2 K$ and $L$ is the film length, we obtain
an expression for
$\eta=(e d R^\square)^{-1} \sqrt{3 \pi /2k_{B} T_{c}
v_{F} l_{i} \hbar}$, which when substituted
into \Eref{Eliashberg17} leads to its good agreement
with both experimental values
of the equilibrium pair-breaking current
and those calculated in the Ginzburg -- Landau
theory, $I_{c}^{GL}(T)$ (see \Eref{Eliashberg19})
\cite{DmitrievZolochevskiiBezuglyi1}.
This expression for the parameter $\eta$ will be used
below in the formula \eref{Eliashberg16}
for enhanced critical current in its comparison
with experimental results.

It is interesting to note that prior to reference
\cite{DmitrievZolochevskiiBezuglyi1}, to
our knowledge, the temperature
dependencies of the enhanced critical current,
arising from \Eref{Eliashberg16}, were
not compared directly with experimental data for $I_{c}^{P}(T)$.
We point out, however, that
attempts to compare, at least qualitatively, experimental
dependencies $I_{c}^{P}(T)$ with
the Eliashberg theory have been made.
For example, in reference \cite{KlapwijkBerghMooij} the
authors presented the pair-breaking
current of Ginzburg -- Landau, using \Eref{Eliashberg18}
for the equilibrium gap, in the form
\begin{eqnarray} \label{Eliashberg19}
I_{c}^{GL}(T)= \frac{c\Phi_{0} w}{6\sqrt{3}\pi^2\xi(0)
\lambda_{\perp}(0)}(1- T/T_{c})^{3/2}=K_{1}
\Delta_{0}^3(T),
\end{eqnarray}
where $\Phi_{0}=\hbar c/2e$ is the magnetic flux quantum.
Since the temperature dependence of the
enhanced critical current in a narrow channel appeared to
be close to equilibrium in its shape,
$I_{c}^{P}(T)\varpropto (1-T/T_{c}^{P})^{3/2}$,
where $T_{c}^{P}$ is the superconducting transition
temperature in a microwave field, the $I_{c}^{P}(T)$
was approximated by an expression similar
to \Eref{Eliashberg19}
\begin{eqnarray} \label{Eliashberg20}
I_{c}^{P}(T)=K_{2} \Delta_{P}^{3}(T).
\end{eqnarray}
where the stimulated energy gap $\Delta_{P}(T)$
was calculated in the Eliashberg theory at zero
superconducting current (\Eref{Eliashberg12} at $j_{s}=0$).
After that, assuming
$K_{1}=K_{2}$ and using the value of the microwave power as
a fitting parameter, the authors
of reference \cite{KlapwijkBerghMooij} fitted calculated values of
$I_{c}^{P}(T)$ to experimental data with a certain degree of accuracy.

It is clear that such a comparison of experimental results
with the Eliashberg theory is only a
qualitative approximation, and cannot be used to
obtain quantitative
results \cite{DmitrievZolochevskiiBezuglyi1}. First of all,
\Eref{Eliashberg19} and
\Eref{Eliashberg20} contain a value of the gap in a
zero-current regime ($j_{s} = 0$), which
differs from that when there is a current. Second,
the pair-breaking curves $j_{s}(\Delta)$ in
the equilibrium state (P = 0) and with a microwave
field are very different
\cite{DmitrievKhristenko2}. Finally, as shown in references \cite{IvlevLisitsyn,KlapwijkBerghMooij,DmitrievKhristenko2} for
$T\rightarrow T_{c}^{P}-0$  the enhanced order parameter
$\Delta_{P}(T)$  tends to a finite
(but small) value $\Delta_{P}(T_{c}^{P})=(1/2)\hbar\omega$
and abruptly vanishes at $T>T_{c}^{P}$,
while the critical current vanishes continuously (without a jump)
at $T \rightarrow T_{c}^{P}$  and,
therefore, cannot in general be satisfactorily described
by the formula of the type \eref{Eliashberg20}.
This is evidenced by a marked deviation of the dependence
\eref{Eliashberg20} from experimental points
in the close vicinity of $T_{c}$. In this review, an analysis
of experimental data is based on
the exact formula \eref{Eliashberg16} with using the
numerical solution of \Eref{Eliashberg15}.
\section{\label{sec:Enhanc}Enhancement of superconductivity
by an external microwave irradiation in films of different widths}
Upon enhancement of superconductivity by an electromagnetic
field the superconducting gap $\Delta$
is, strictly speaking, variable in space and time.
However, for sufficiently thin and narrow
superconducting samples the dependence of $\Delta$ on
coordinates can be neglected. Moreover, near $T_{c}$
the relaxation time of the order parameter
$\tau_{\Delta}\simeq1.2\tau_{\varepsilon}/(1-T/T_{c})^{1/2}$
is large compared to the inverse frequency of stimulating
microwave radiation ($\omega\tau_{\Delta}\gg1$), and the
temporal oscillations of $\Delta$ can also be neglected.
Therefore, the microscopic theory of
Eliashberg \cite{Eliashberg1} included neither the time
nor spatial variations of the order parameter in a sample.

To realize experimentally the case discussed in
this theory of a spatially homogeneous
non-equilibrium state of a superconductor in a
high-frequency field it was necessary to ensure
the constancy of the energy gap over the sample
volume ($w, d \sim \xi(T), \lambda_{\bot}(T)$).
This is a purely technological problem.
It was also important to ensure a uniform distribution
of the transport current through the sample volume.
Failure to do so resulted in a
non-uniform distribution of $\Delta$ due not to
technological reasons, but because of
the dependence of the energy gap on the transport current
$\Delta(I)$. Finally, it was important
to provide an efficient heat transfer from a sample.
It was shown that narrow superconducting
film channels deposited on relevant substrates
best met these requirements. The theory proposed
in \cite{Eliashberg1} has been fully confirmed
in experimental studies of such samples (see,
e.g., references \cite{DmitrievKhristenko} and
\cite{DmitrievGubankovNad}).

For wider films, electro dynamical changes
of $\Delta$ over the film
width with a non-uniformly distributed
current and with the presence of intrinsic vortices cannot be
neglected. Therefore, the theory \cite{Eliashberg1},
strictly speaking, does not apply in the
case of wide films. Yet, although it is difficult
to develop a theory in the case of a
non-uniform distribution of $\Delta$ in a superconductor,
in principle there should also be
an effect of enhancement of superconductivity in this case.

\begin{figure}
\centerline{\includegraphics[height=3in]{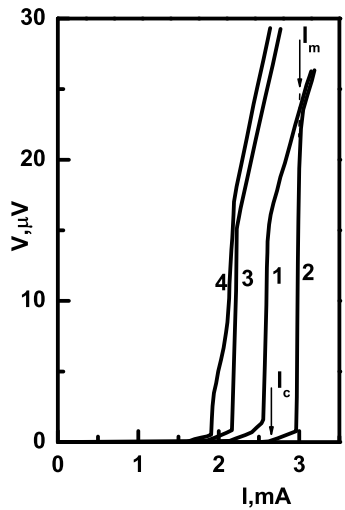}}
\caption{
%Fig.1.
A family of current-voltage characteristics of the
film sample SnW5 at $T = 3.745$~K and $f = 12.89$~GHz
for various levels of irradiation power: the
irradiation power is zero (1), with increasing a serial
number of the IVC the irradiation power increases (2-4).
} \label{f}
\end{figure}

In 2001, the enhancement of superconductivity by an
external electromagnetic field has also
been found in wide, $w \gg \xi(T), \lambda_{\bot}(T)$,
high-quality superconducting films of
tin \cite{AgafonovDmitriev} with a non-uniform spatial
distribution of $\Delta(I)$ over the
sample width. It has been experimentally shown that
under an external electromagnetic field not
only the critical current $I_{c}$ increases but also
the current of formation of the
first phase-slip line (see \fref{f}) does so.
In reference \cite{AgafonovDmitriev}, this current
is designated as $I_{c}^{dp}$.

In reference \cite{DmitrievZolochevskiiSalenkova2} the
temperature dependencies of the current $I_{c}^{dp}$ was
analyzed with taking into account the nontrivial distribution
of the transport current and the density of vortices over
a wide film. As a result, it was shown that the current
$I_{c}^{dp}$ is the critical pair-breaking current of
Ginzburg -- Landau $I_{c}^{GL}$, if a film corresponds
to the parameters of a vortex-free narrow channel in the
temperature region near $T_{c}$. Far from $T_{c}$
this current is the maximum current of existence of the
vortex resistive state $I_{m}$ in
the Aslamazov -- Lempitsky theory \cite{AslamazovLempitsky}.
A physical meaning of the current $I_{m}$ is that it is
the maximum current, at which a steady uniform flow of
intrinsic vortices of the transport current across a
wide film is still possible. If
it is exceeded, $I>I_{m}$, the vortex structure collapses,
and in its place there appears
a structure of phase-slip lines \cite{DmitrievZolochevskii}.
It is the phase that determines
the resistivity of a sample for a further increase in the
transport current.

In this connection, the problem of superconductivity
enhancement in wide films becomes
particularly interesting, because it requires consideration
of the behavior in a microwave field
not only of both the critical current and the critical
temperature. An important object of the
investigation is the current $I_{m}$, as well as its
relation with $I_{c}$ under an external
electromagnetic irradiation of different
frequencies $f$ and power $P$.

\begin{table}
\caption{Parameters of tin film samples: $L$ is the length, $w$ is
the width, $d$ is the thickness of the sample, $l_{i}$ is the
electron mean free path, and $R^\square=R_{4.2}w/L$ } \label{tab}
%\begin{indented}
\begin{center}
\begin{tabular}{lllllllll}
\br Sample &    $L$, &   $w$,  &   $d$, &  $R_{4.2}$, & $R^\square$, &
$T_{c}$, & $l_{i}$, &  $R_{300}$, \\
       &  $\mu$m &  $\mu$m &   nm   &  $\Omega$   &  $\Omega$    &
K       &  nm      &  $\Omega$   \\
\mr Sn1   &  64     &   1.5    &    90 &  3.05      &  0.071 &
3.834  &   174   &   59 \\

SnW5  &   92     &   42    &    120 &  0.14      &  0.064       &
3.789  &   145   &   2.270 \\

SnW6  &   81     &   17    &    209 &  0.185      &  0.039       &
3.712  &   152   &   3.147 \\

SnW8  &   84     &   25    &    136 &  0.206      &  0.061       &
3.816  &   148   &   3.425 \\

SnW10  &   88     &   7    &    181 &  0.487      &  0.040       &
3.809  &   169   &   9.156 \\
\br
\end{tabular}
%\end{indented}
\end{center}
\end{table}
This section presents the results of studying a dependence
of enhancement of the critical
current $I_{c}$ and the current of formation of the first
PSL $I_{m}$ on power and frequency of
an electromagnetic field in thin
(thickness $d \ll \xi(T), \lambda_{\bot}(T)$)
superconducting films as a function of their width
$w$ \cite{DmitrievZolochevskiiSalenkova2}. To understand
how the effect of superconductivity enhancement manifests
itself in a wide film, the authors of this work increased
gradually the sample width starting from a narrow channel,
and observed how the phenomenon of
superconductivity enhancement was changing. As samples
thin ($d \ll \xi(T), \lambda_{\bot}(T)$) tin films
which were prepared as described in reference
\cite{DmitrievZolochevskii} were used. This
original technology enabled to
minimize defects in both the film edge and
its volume. The critical current of these samples is
determined by suppression of the barrier for
entering vortices when the current density at the edge
of the film is of the order of $j_{c}^{GL}$, and
reaches the maximum possible theoretical
value \cite{AslamazovLempitsky}, indicating
the absence of edge defects which create local lowering
of the barrier and thus reduce the $I_{c}$.
IVC were measured by a four-probe method. In measuring
the IVC samples were placed in a double screen
of annealed permalloy. In the area of the sample
the magnetic field was:
$H_{\bot}= 7\times10^{-4}$ Oe,
$H_{\|}= 6,5\times10^{-3}$ Oe. To supply
an electromagnetic irradiation to the film
sample it was placed in a rectangular wave-guide parallel
to the electric field component in the wave-guide,
or irradiated from the shorted end of a coaxial line,
or the sample was connected to a 50 - ohm coaxial
line through a separating capacitance (contact method).
The parameters of the samples are shown in Table 1.
The temperature was measured by the vapor pressure
using mercury and oil pressure gauges. In doing so,
an influence of the microwave field introduced
into the cryostat during the experiment on
measurements of temperature in the case of
using electronic thermometers was excluded.
The temperature stabilization (helium vapor
pressure) was provided by a membrane
manostat with accuracy better than $10^{-4}$ K.

In reference \cite{AgafonovDmitriev} it was shown
that long ($L\gg \xi(T), \lambda_{\bot}(T)$) and wide
($w \gg \xi(T), \lambda_{\bot}(T)$) superconducting
films reveal the effect of increasing the critical
current and the current $I_{m}$ under an external
microwave irradiation. \Fref{f} shows a family of
current-voltage characteristics of one of these films
(SnW5) of the width of 42 $\mu$m for different
power levels of microwave irradiation with a
frequency $f=12.89$ GHz.
Here, as in reference \cite{DmitrievZolochevskii} the
notation are introduced: $I_{c}(T)$ is the current of
voltage appearance across
the sample as a result of entering vortices of its
intrinsic magnetic flux current, $I_{m}(T)$ is the
maximum current of existence of a stable uniform
flow of intrinsic vortices or the current of formation
of the first phase-slip line. In \fref{f} it is seen
that the current $I_{c}(P)$ (see \fref{f},
IVC 2) is significantly higher than $I_{c}(P = 0)$
(see \fref{f}, IVC 1), and $I_{m}(P)>I_{m}(P=0)$.
Thus, under external irradiation both $I_{c}$ and
$I_{m}$ increase \cite{AgafonovDmitriev}.
\begin{figure}
\centerline{\includegraphics[height=3in]{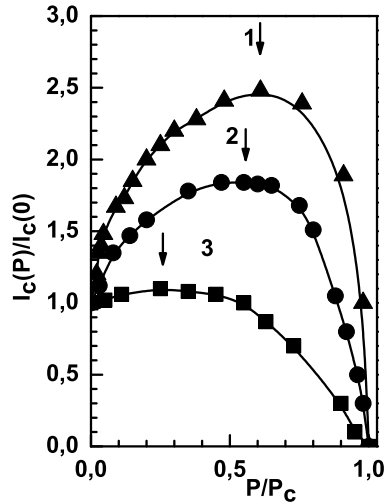}}
\caption{
%Fig.2.
The dependence of the relative critical current
$I_{c}(P)/I_{c}(0)$ in the sample Sn1 on the reduced
microwave irradiation power $P/P_{c}$ at $T= 3.812$~K
for different irradiation frequencies $f$, GHz:
15.4~($\blacktriangle$), 8.1~(\fullcircle),
3.7~(\fullsquare) ($I_{c}(0)$ is the critical current of the
film at $P=0$; $P_{c}$ is the minimum power of
electromagnetic irradiation at which $I_{c}(P)=0$).
} \label{f2}
\end{figure}
\subsection{\label{sec:CRICUR1}The critical current}
For a narrow channel Sn1 of the width of $w = 1.5~\mu$m
at $T/T_{c} = 0.994$ and
$w/\lambda_{\bot}(T~=~3.812~K)~=~0.28$ as a
function of the reduced power $P/P_{c}$ of microwave irradiation
the relative magnitude of the effect of enhancement of the
critical superconducting current
$I_{c}(P)/I_{c}(0)$ is shown in \fref{f2}
for various frequencies of an external irradiation.
Here, $P_{c}$ is the minimum power at which
$I_{c}(P=P_{c})=0$. The curve 3 corresponds to a
low enough frequency of irradiation, 3.7~GHz, the
curve 2 is plotted for the irradiation frequency of
8.1~GHz; the curve 1 corresponds to the
frequency of 15.4~GHz. The arrows indicate the values
of powers under which the maximal effect of
enhancement $I_{c}$ was observed for each of the
irradiation frequencies. For the irradiation
frequency $f$=3.7~GHz the reduced power of microwave
irradiation, at which a maximum of the effect
is observed, equals $P/P_{c}$=0.25. For the frequency
$f$=8.1~GHz, $P/P_{c}$=0.51, and for
$f$=15.4~GHz, $P/P_{c}$=0.61. It is seen that with
increasing the irradiation frequency the reduced
power, at which a maximum of the enhancement effect is
observed, increases
\cite{DmitrievZolochevskiiSalenkova2}. Unfortunately, in
the enhancement theory \cite{Eliashberg1}
a reduction of the effect, after the maximum, with
increasing the irradiation power is not considered.
Therefore, a shift of maximal manifestation of
superconductivity enhancement under electromagnetic
irradiation towards higher power with increasing the
frequency the theory \cite{Eliashberg1} cannot explain.

\begin{figure}
\begin{center}
\includegraphics[height=3in]{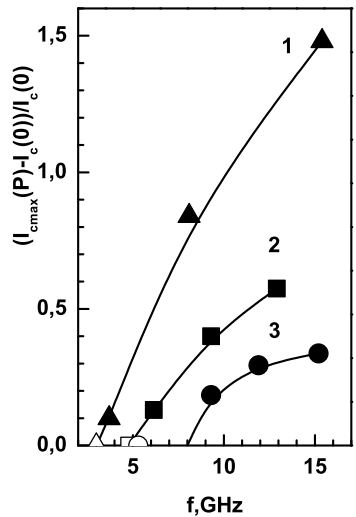}
\end{center}
\caption{\label{f3}
The reduced value of exceeding the maximum critical
current $I_{cmax}( P)$ over $I_{c}(0)$ as a
function of the irradiation frequency for the samples
Sn1~($\blacktriangle$), SnW10~(\fullsquare)
and SnW5~(\fullcircle) at $T/T_{c}\approx$ 0.99;
the values of lower cut-off frequencies of
superconductivity enhancement calculated by
\Eref{Eliashberg11} for the samples Sn1~(\opentriangle),
SnW10~(\opensquare)) and SnW5~(\opencircle).}
%Fig.3.
\end{figure}

The reduced excess of the critical current as a function
of the irradiation frequency for films of
different width is shown in \fref{f3}.
It is seen that with increasing the frequency the
effect of exceeding the critical current $I_{cmax}(P)$
over $I_{c}(P=0)$ increases for both narrow (curves 1 and
2) and wide (curve 3) films. With
further increase of the frequency this dependence passes
through a maximum and then begins to decrease
(not shown here). It should be noted that the frequency
at which the maximum effect of enhancement of the
critical current is observed, decreases with
increasing the film width (for Sn1 the maximum frequency is
about 30~GHz, and for SnW5 about 15~GHz).
It is interesting to note that for the film Sn1
(see \fref{f3}, curve 1) the calculation of the
lower cut-off frequency of enhancement $f_{L}$
from \Eref{Eliashberg11} gives the value of 3~GHz
(indicated in \fref{f3} by a symbol~
\opentriangle ),
which corresponds well to the experiment, as was
shown previously for narrow channels \cite{DmitrievKhristenko}.
It is important to emphasize that for
the calculation of the lower cut-off frequency of
enhancement for the sample Sn1, the value
$\tau_{\varepsilon} = 8.3\times10^{-10}$ s typical
of this series of samples was used.

\begin{figure}
\begin{center}
\includegraphics[height=3in]{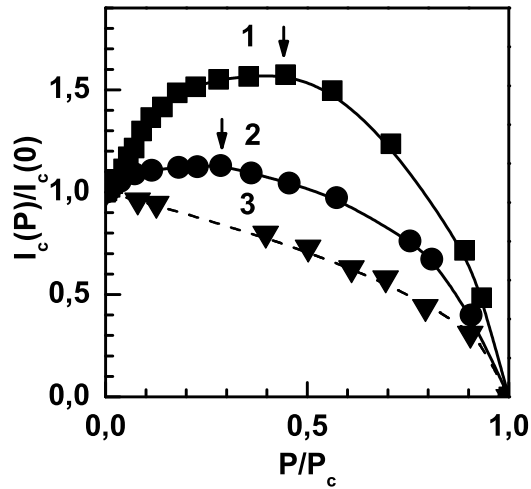}
\end{center}
\caption{\label{f4}
The dependence of the relative critical current
$I_{c}(P)/I_{c}(0)$ in the sample SnW10 on the
reduced microwave irradiation power $P/P_{c}$ at
$T= 3.777$~K for different irradiation frequencies
$f$, GHz: 12.91~(\fullsquare), 6.15~(\fullcircle),
0.63~($\blacktriangledown$); dashed curve 3
is the dependence $I_{c}(P)/I_{c}(0)(P/P_{c})$
calculated by \Eref{Eliashberg21}.}
%Fig.4.
\end{figure}

The dependencies of the reduced critical current
$I_{c}(P)/ I_{c}(0)$ on the reduced power $P/P_{c}$ of a
microwave field for different irradiation frequencies
for the widest sample SnW10 of the width of
7 $\mu$m at a temperature $T~=~3.777$~K
($T/T_{c}$~=~0.992) are shown in \fref{f4}.
At this temperature $w/\lambda_{\bot}$~=~3.56,
i.e., less than 4. As shown in references
\cite{DmitrievZolochevskii} and
\cite{DmitrievZolochevskiiBezuglyi}, at this
temperature, the sample SnW10 is still a narrow
channel and there is no resistive part, caused
by the motion of Pearl
vortices, in its current-voltage
characteristics. Indeed, the dependencies
1 and 2 in \fref{f4} do not
differ qualitatively from those curves in
\fref{f2}. The arrows in \fref{f4} have
the same meaning as in \fref{f2}. In \fref{f4}
it is seen that on increasing the irradiation
frequency the reduced power
at which the maximum effect of
superconductivity enhancement is observed,
increases as it was for a narrow
channel. Moreover, the calculation of the
lower cut-off frequency from \Eref{Eliashberg11}
gives the value of 4.8~GHz (denoted by a symbol~
\opensquare), which also agrees quite well with the experimental
value $f_{L}$, as seen in \fref{f3} (curve 2). It is
important to note that to calculate the lower
cut-off frequency of enhancement for the sample SnW10, the value
$\tau_{\varepsilon}=4.3\times10^{-10}$ s
typical of this series of samples was used.

In \fref{f4}, the experimental dependence
($\blacktriangledown$) was obtained for the relatively
low radiation frequency ($f=0.63$ GHz). This
frequency is below the cut-off frequency of the effect of
superconductivity enhancement, $f_{L}$, so there is only
a suppression of $I_{c}$ with increasing $P$.
Since in these experimental conditions the sample
SnW10 is a narrow channel, it is interesting to compare
the experimental dependence ($\blacktriangledown$)
and the theoretical curve 3. In reference
\cite{BezuglyiDmitrievSvetlov} it is shown
that for superconducting films, the critical current
of which is equal to the pair-breaking current
of Ginzburg -- Landau, the following dependence of the
critical current on the irradiation power of
electromagnetic field is valid:
\begin{eqnarray}\label{Eliashberg21}
I_{c}(P, \omega)/I_{c}(T)=[1-(P/P_{c}(\omega))]^{1/2} \\   \nonumber
\times [1-(2P/(\omega\tau_{\Delta})^{2}P_{c}(\omega))]^{1/2}
\end{eqnarray}
at $\omega \tau_{\Delta}\gg 1$. In our case
$\omega \tau_{\Delta}\approx$24 and the calculated
dependence \eref{Eliashberg21} is shown in \fref{f4}
by a dashed curve 3. It is seen that it agrees quite
well with the experimental dependence ($\blacktriangledown$)
and confirms the conclusion made in
reference \cite{DmitrievZolochevskii} that at
$w/\lambda_{\bot}~<~4$ films are narrow channels.
On lowering the temperature of the sample SnW10
below the crossover temperature $T_{cros1}$
\cite{DmitrievZolochevskii}, the relation
$w/\lambda_{\bot}$ increases and becomes slightly greater than 4.
This is due to a gradual decrease of
$\lambda_{\bot}(T)$ upon changing the temperature far
away from $T_{c}$. As a result, the distribution of
the transport current over the width of the film
becomes non-uniform, but not enough to significantly
affect the behavior of the film in an electromagnetic field,
and consequently the form of $I_{c}(P)$.
To observe significant differences it is necessary to
lower significantly the temperature, but the effect
of superconductivity enhancement decreases markedly
in this case. This is due to a decrease in the number
of excited quasiparticles above the gap
\cite{Eliashberg1,DmitrievKhristenko,DmitrievGubankovNad}.
\begin{figure}
\begin{center}
\includegraphics[height=3in]{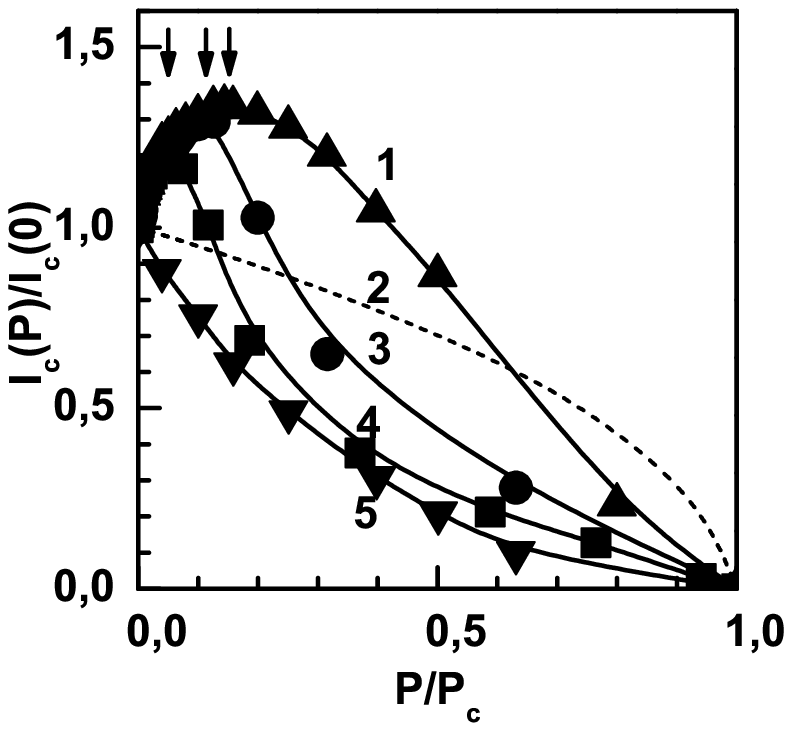}
\end{center}
\caption{\label{f5}
The dependence of the relative critical current
$I_{c}(P)/I_{c}(0)$ in the sample SnW5 on
the reduced microwave radiation power $P/P_{c}$ at
$T = 3.744$~K for different irradiation frequencies
$f$, GHz: 15.2 ($\blacktriangle$), 11.9~(\fullcircle),
9.2~(\fullsquare), 5.6~($\blacktriangledown$).
The dashed curve 2 is the dependence
($I_{c}(P)/I_{c}(0))(P/P_{c}$) calculated by
\Eref{Eliashberg21}.}
%Fig.5.
\end{figure}
Therefore, to further investigate the effect of
superconductivity enhancement an initially wider film
(the SnW5 sample of the width of 42 $\mu$m) should be taken.
In \fref{f5} for this sample
at $T/T_{c}$~=~0.988 and $w/\lambda_{\bot}(T~=~3.744$~K)~=~20
there are dependencies of the
reduced critical current $I_{c}(P)/I_{c}(0)$ on the reduced
power $P/P_{c}$ of a microwave field
with different irradiation frequencies.
The meaning of the arrows is the same as in
\fref{f2} and \fref{f4}. A \fref{f5} shows that the
reduced power, at which the maximum effect of
superconductivity enhancement is observed, increases
with the irradiation frequency
\cite{DmitrievZolochevskiiSalenkova2}. Moreover, it is seen that
descending parts of the dependencies 1, 3--5 in \fref{f5}
differ from those in \fref{f2} and
\fref{f4} by a curvature sign: in \fref{f2} and \fref{f4},
descending parts of the curves are convex,
while in \fref{f5} they are concave. The curve 5 was
obtained at the irradiation frequency
$f$= 5.6~GHz, and in this case the enhancement
effect was not observed. The dotted curve 2 shows the
calculated dependence $I_{c}(P)$ by \Eref{Eliashberg21}
for the SnW5 film if the transport current in
it was distributed uniformly over its width. It
is seen that the curves 2 and 5 are significantly different
from each other. Therefore, the concavity of the descending
part of the experimental dependence 5 may
well be attributed to the non-uniform current
distribution across the width of the film.
\begin{figure}
\centerline{\includegraphics[height=3in]{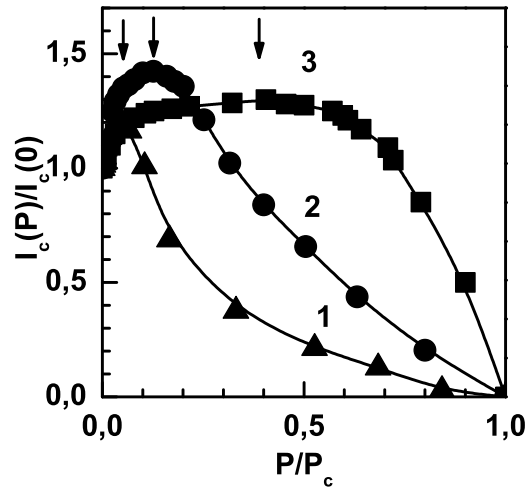}}
\caption{
%Fig.6.
The dependence of the relative critical current
$I_{c}(P)/I_{c}(0)$ on the reduced microwave irradiation
power $P/P_{c}$ with the frequency $f$= 9.2~GHz at
$T/T_{c} \approx$ 0.99 in different samples:
SnW5~($\blacktriangle$), SnW6~(\fullcircle) and
SnW10~(\fullsquare).
} \label{f6}
\end{figure}
In \fref{f5} the dependencies 1, 3 and 4
were obtained for irradiation frequencies: 15.2, 11.9,
and 9.2~GHz, respectively. The concavity of descending
parts can be associated, as for the curve 5, with
non-uniform current distribution over the sample width.

Interestingly, in the narrow film Sn1 the enhancement
effect is already clearly seen at the irradiation
frequency $f$ = 3.7~GHz (see \fref{f2}, curve 3),
while in the film SnW5 it is not observed even
at $f$= 5.6~GHz (see \fref{f5}, curve 5).
The calculation of $f_{L}$ for the film by the formula
\eref{Eliashberg11} gives the value of 5.1~GHz,
which no longer corresponds to the experimental value of 8.0~GHz.
It is important to emphasize that for calculation of the
lower cut-off frequency of enhancement in the
sample SnW5, the value $\tau_{\varepsilon}~=~4~\times~10^{-10}$~s
typical of this series of samples was used.

\begin{figure}
\begin{center}
\includegraphics[height=3in]{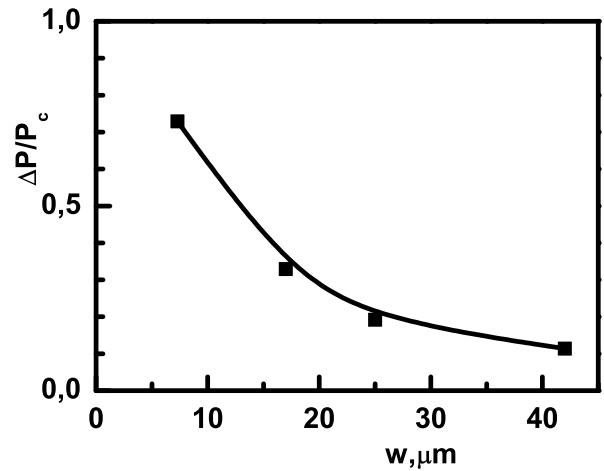}
\end{center}
\caption{\label{f7}
A region of external irradiation power $P/P_{c}$,
where the effect of enhancement of the critical
current is observed, as a function of the film
width $w$ for the frequency
of 9.2~GHz at $T/T_{c} \approx$ 0.99.}
%Fig.7.
\end{figure}

The dependencies $I_{c}(P)$ in relative units for films
of different widths for the same experimental
conditions are shown in \fref{f6}. The arrows in
\fref{f6} have the same meaning as in
\fref{f2} and \fref{f4}. In \fref{f6} it is
seen that with growth of the film width the
ratio $P/P_{c}$, at which there is a maximum
enhancement effect, is reduced, and the effect of
enhancement of the critical current in wider films
is observed at lower radiation power, since in the
microwave range the value of $P_{c}$ is practically
independent of frequency
\cite{PalsRamekers, BezuglyiDmitrievSvetlov}.
\Fref{f7} shows the dependence of power region of
external irradiation, $P/P_{c}$, where the effect
of enhancement of the critical current is
observed, on the film width $w$ at a fixed irradiation
frequency and temperature. From the data in
the figure it follows that as the film width increases
the power range $\Delta P$, where the effect
of superconductivity enhancement is observed, is reduced.
Therefore, one can assume that for rather wide
tin films ($w > 100~\mu$m) the effect of superconductivity
enhancement can be practically
unrealisable in experiment as due to a very narrow power
range of existence of this effect and
because of its small size.
\subsection{\label{sec:MAXCUR1}A maximum current
of the existence of vortex resistivity}
In subsection~\ref{sec:CRICUR1} we have found
out how electromagnetic irradiation affects the
critical current $I_{c}$ of films of different widths.
Another important characteristic current of a wide
film is the so-called maximum
current of the existence of vortex resistivity
$I_{m}$. Experimentally the current $I_{m}$ was studied in
reference \cite{DmitrievZolochevskii} and has the
form \cite{AslamazovLempitsky}:
\begin{eqnarray}\label{Eliashberg22}
I_{m}(T)=CI_{c}^{GL}(T)\ln^{(-1/2)}(2w/\lambda_{\bot}(T))
\end{eqnarray}
To date there is no theory of superconductivity
enhancement in wide films, and therefore, at present the results
of experimental studies of enhancement of $I_{m}(T)$
cannot be compared with theoretical predictions.
However, from \Eref{Eliashberg22} obtained for the
equilibrium (without external irradiation)
current $I_{m}(T)$, it can be assumed that the
behavior of $I_{m}(P,f)$ in an electromagnetic field
is determined by the effect of the irradiation on
$I_{c}^{GL}(T)$ and $\lambda_{\bot}(T)$.

Upon enhancement of superconductivity $I_{c}^{GL}(T)$
increases, and $\lambda_{\bot}(T)$,
according to general considerations (enhancement
of $T_{c}$), must decrease. The reduction rate of
$\lambda_{\bot}(T)$ also depends on the proximity
of the operating temperature $T$ to $T_{c}$,
other things being equal. Thus, it is clear qualitatively
that the rate of increasing $I_{m}(P)$
must be lower than the growth rate of $I_{c}(P)$.
\begin{figure}
\centerline{\includegraphics[height=3in]{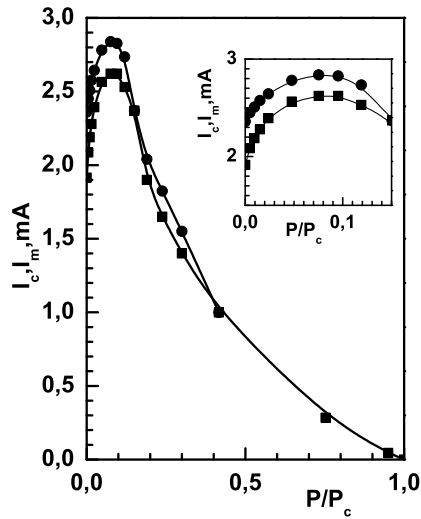}}
\caption{
%Fig.8.
The dependence of the critical current $I_{c}$
(\fullsquare) and the maximum current of existence of
the vortex resistivity $I_{m}$ (\fullcircle) for the
sample SnW5 on the reduced microwave power $P/P_{c}$
with the frequency $f$=12.89~GHz at $T = 3.748$~K.
The inset shows an enlarged fragment of the
above-mentioned dependencies.
} \label{f8}
\end{figure}
\Fref{f8} shows the experimental dependencies
$I_{c}(P)$ and $I_{m}(P)$ for the film SnW5
\cite{DmitrievZolochevskii23}. The inset shows the
initial parts of the curves for a more visual
representation of the growth rate of $I_{c}(P)$ and
$I_{m}(P)$. In the figure it is seen that indeed with
increasing the irradiation power of the film its
critical current $I_{c}(P)$ increases faster than the
current $I_{m}(P)$. The question arises whether there
is enough change in $\lambda_{\bot}(T)$ under
electromagnetic irradiation to suppress the growth
of $I_{m}(P)$ in comparison with an increase
of $I_{c}(P)$. An estimation of changes in
$\lambda_{\bot}(P)$ at the relative temperature at which the
curve in \fref{f8} was measured, indicates that they
are sufficiently small, and in accordance
with \Eref{Eliashberg22} they cannot slow down
significantly an increase of $I_{m}(P)$. Therefore,
there must be another reason. An analysis of
experimental data suggests that the reason may be the
non-uniformity of the current distribution over the
film width and the presence of resistive vortex
background, which affects $I_{m}$. This background
also depends on an external irradiation, what
\Eref{Eliashberg22} for the equilibrium current
$I_{m}$ ignores. In this context, it is necessary to
draw attention to the fundamental difference between
$I_{c}$ and $I_{m}$. $I_{c}$ always appears against a
background of the pure superconducting state. So due
to the transverse Meissner effect it is always
first achieved at the film edges. And the larger
its width with respect to $\lambda_{\bot}$, the more
inhomogeneous distribution of the transport current
is in it. In contrast, the current $I_{m}$ is
the maximum current at which a uniform flow of
vortices across the film is still possible. The presence
of a moving vortex lattice makes the distribution
of the superconducting current across the film more
uniform, although specific \cite{AslamazovLempitsky}.
Thus, in a wide film being in a vortex-free
state at $I \leq I_{c}$ the current is always more
non-uniformly distributed over the width than in the
same film in the presence of intrinsic vortices for
the currents $I_{c} < I \leq I_{m}$. Because
of the above reasons, there is a need of new
theory of non-equilibrium state of a wide film, which
could take into account the non-uniform distribution
of the transport current and the order
parameter over the film width in calculating
$I_{c}(P, f)$ and the presence of the vortex
resistivity $R(P, f)$ when calculating $I_{m}(P, f)$.
\section{\label{sec:Temp}Temperature dependencies of
currents enhanced by microwave power in wide films}
\subsection{\label{sec:MAXCUR2}The critical current}
This section presents results of systematic study
of the critical current enhancement in wide
superconducting films. It is established that the
main properties of superconductivity enhancement in
wide films with non-uniform current distribution over
the cross section of a sample and those in
narrow channels are very similar
\cite{DmitrievZolochevskiiBezuglyi1}.
A relative moderation of the current non-uniformity
in wide films near $T_{c}$ allowed for using,
with a little change, the theory of superconductivity
enhancement in spatially homogeneous systems
to interpret experimental results in wide films.

\begin{figure}
\centerline{\includegraphics[height=3in]{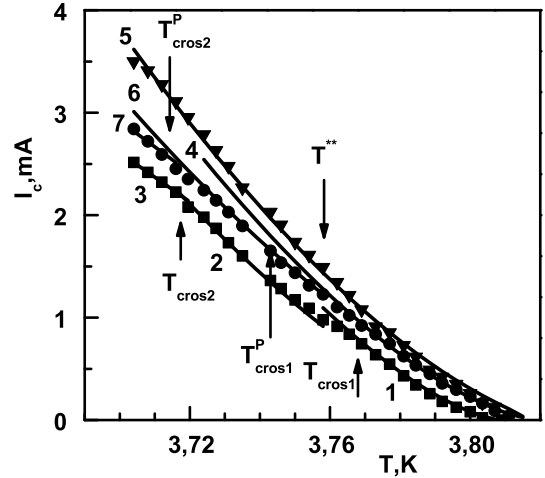}}
\caption{
%Fig.9.
The experimental temperature dependence of the
critical currents $I_{c}(P=0)$~(\fullsquare),
$I_{c}(f = 9.2~GHz)$~(\fullcircle), and
$I_{c}(f = 12.9~GHz)$~($\blacktriangledown$) for the sample SnW10.
The theoretical dependence
$I_{c}^{GL}(T)=7.07\times 10^{2}(1 - T/T_{c})^{3/2}$~mA
calculated by \Eref{Eliashberg19}
\cite{DmitrievZolochevskiiSalenkovai25} (curve 1);
calculated dependence $I_{c}(T)=5.9\times 10^{2}(1 - T/T_{c})^{3/2}$~mA
(curve 2);
theoretical dependence
$I_{c}^{AL}(T) =9.12\times 10^{1}(1 - T/T_{c})$~mA
calculated by
\Eref{Eliashberg27} \cite{AslamazovLempitsky}~(straight line 3);
theoretical dependence $I_{c}(f = 9.2~GHz)$
calculated by \Eref{Eliashberg16} and
fitting dependence
$I_{c}(T)=6.5\times 10^{2}(1 - T/3.818)^{3/2}$~mA
(curve 4);
theoretical dependence $I_{c}(f=12.9~GHz)$ calculated
by \Eref{Eliashberg16}, and fitting
dependence
$I_{c}(T)=6.7\times 10^{2} (1 - T/3.822)^{3/2}$~mA
(curve 5);
theoretical dependence $I_{c}(f = 9.2~GHz)$
calculated by \Eref{Eliashberg16} normalized by the
curve 2, and fitting dependence
$I_{c}(T)=5.9\times 10^{2}(1 - T/3.818)^{3/2}$~mA
(curve 6);
calculated dependence
$I_{c}( T)=9.4\times 10^{1}(1 - T/3.818)$~mA
(straight line 7).
} \label{f9}
\end{figure}

\Fref{f9} shows experimental temperature dependencies
of the critical current for the sample SnW10
\cite{DmitrievZolochevskiiBezuglyi1}. At first, a
behavior of $I_{c}(T)$ without an external
electromagnetic field (see \fref{f9},~($\fullsquare$))
is considered. A width of the film SnW10 is
relatively small ($w$=7 $\mu$m), so in the temperature
range $T_{cros1}<T<T_{c}=3.809$~K close enough
to $T_{c}$, the sample behaves like a narrow channel,
and the critical current is equal to the
pair-breaking current of Ginzburg -- Landau
$I_{c}^{GL}(T)\varpropto (1-T/T_{c})^{3/2}$ which indicates
the high quality of the sample. The crossover temperature
$T_{cros1}=3.769$~K corresponds to the
transition of the sample in the wide film regime:
at $T < T_{cros1}$ there is a vortex part in the IVC.
The temperature dependence $I_{c}(T)$ at $T < T_{cros1}$
initially retains the form
$(1 - T/T_{c})^{3/2}$, although the value of $I_{c}(T)$
turns out to be less than the pair-breaking
current $I_{c}^{GL}(T)$  due to the appearance of a
non-uniform distribution of the current density and
its decrease far away from the film edges. Finally,
when $T < T_{cros2}=3.717$~K the temperature
dependence of the critical current becomes linear
$I_{c}(T)=I_{c}^{AL}(T)= 9.12\times 10^{1}(1 - T/T_{c})$~mA,
which corresponds to the
Aslamazov--Lempitsky theory \cite{AslamazovLempitsky}.
The latter fact confirms our earlier
conclusion about the high quality of the film sample SnW10.

For measurements in a microwave field
\cite{DmitrievZolochevskiiBezuglyi1} the irradiation power was
chosen in such a way that the critical current $I_{c}^{P}(T)$
was maximal. Consider the behavior of
$I_{c}^{P}(T)$ in the sample SnW10 in a microwave field
with a frequency $f$=9.2~GHz (\fref{f9},~(\fullcircle)). In the
temperature range ($T_{cros1}^{P}(9,2~GHz)<T<T_{c}^{P}(9,2~GHz)$)
($T_{cros1}^{P}(9,2~GHz)=3.744$~K, $T_{c}^{P}(9,2~GHz)=3.818$~K)
there is no vortex part in the IVC, i.e., the sample behaves
as a narrow channel. Note that
$T_{cros1}^{P}(9,2~GHz)<T_{cros1}(P=0)$, whereas $T_{c}<T_{c}^{P}$,
i.e., at optimal enhancement the
narrow channel regime retains in a broader temperature range
than in the equilibrium condition. In
the temperature range   $T^{**}=3.760~K<T<T_{c}^{P}$
the experimental values of $I_{c}^{P}(T)$
(see \fref{f9},~(\fullcircle)) are in good agreement
with those calculated by \Eref{Eliashberg16}
(\fref{f9}, curve 4), in which the microwave power
(the value of $\alpha$) is a fitting
parameter \cite{DmitrievZolochevskiiBezuglyi1}.

However, at $T<T^{**}$ the experimental values of
$I_{c}^{P}(T)$ are lower than the theoretical curve 4.
It should be noted that such a deviation from the
theory was observed in the study of narrow aluminium
film channels \cite{KlapwijkBerghMooij}. Nevertheless,
the experimental points fall well on the curve 6
(\fref{f9}). This curve is calculated by the formula
\eref{Eliashberg16}, normalized by an additional numerical
factor which has provided consent to this formula at zero
microwave field to the equilibrium critical current
$I_{c}(T)$. The authors of reference
\cite{DmitrievZolochevskiiBezuglyi1} consider this as a
form factor, which is understood as an estimate of
non-uniformity of the current distribution over the
film width. Finally, at temperatures
$T<T_{cros2}^{P}(9.2~GHz) =3.717$~K the temperature
dependence of the enhanced critical current becomes
linear (\fref{f9}, straight line 7).

\Fref{f9} also shows the temperature dependence
of the highest enhanced critical current of the
sample SnW10 at a higher irradiation frequency $f$=12.9~GHz (\fref{f9},~($\blacktriangledown$))~\cite{DmitrievZolochevskiiBezuglyi1}.
It can be seen that, as in a narrow
channel, the highest enhanced critical current increases
with the irradiation frequency. Note that at
the given irradiation frequency there is no vortex part
in the IVC over the entire temperature
range investigated (up to temperatures $T = 3.700$~K and
even a bit lower). In other words, in the
temperature range
($T_{cros1}^{P}(12,9~GHz)<T<T_{c}^{P}(12,9~GHz)$)~
$T_{cros1}^{P}(12.9~GHz)<3.700$~K and
it is not shown in \fref{f9}; $T_{c}^{P}(12.9~GHz)=3.822$~K)
the sample behaves as a narrow channel. Note that $T_{cros1}^{P}(12.9~GHz)<T_{cros1}^{P}(9,2~GHz)<T_{cros1}(P=0)$,
whereas $T_{c}< T_{c}^{P}(9.2~GHz)<T_{c}^{P}(12.9~GHz)$.
Thus, in conditions of optimal enhancement of
superconductivity the temperature range where the sample
behaves as a narrow channel increases with
the irradiation frequency \cite{DmitrievZolochevskiiBezuglyi1}.

It is also important to note that the experimental
dependence $I_{c}^{P}(T)$ ($\blacktriangledown$) at $f$= 12.9~GHz is
in good agreement with the theoretical one obtained in
calculating the enhanced critical current by
\Eref{Eliashberg16} for a narrow channel (\fref{f9}, curve 5)
over the entire temperature range, and
is well approximated by the dependence
$I_{c}(T)=6.7\times 10^{2}(1 - T/3.822)^{3/2}$~mA.
Hence, it follows that the temperature   of the transition
to the wide film regime, where the vortex region appears
in the IVC, as well as the deviation temperature $T^{**}$
of the experimental dependence from
the dependence, calculated by \Eref{Eliashberg16},
decrease with increasing the irradiation
frequency \cite{DmitrievZolochevskiiBezuglyi1}.

\begin{figure}
\centerline{\includegraphics[height=3in]{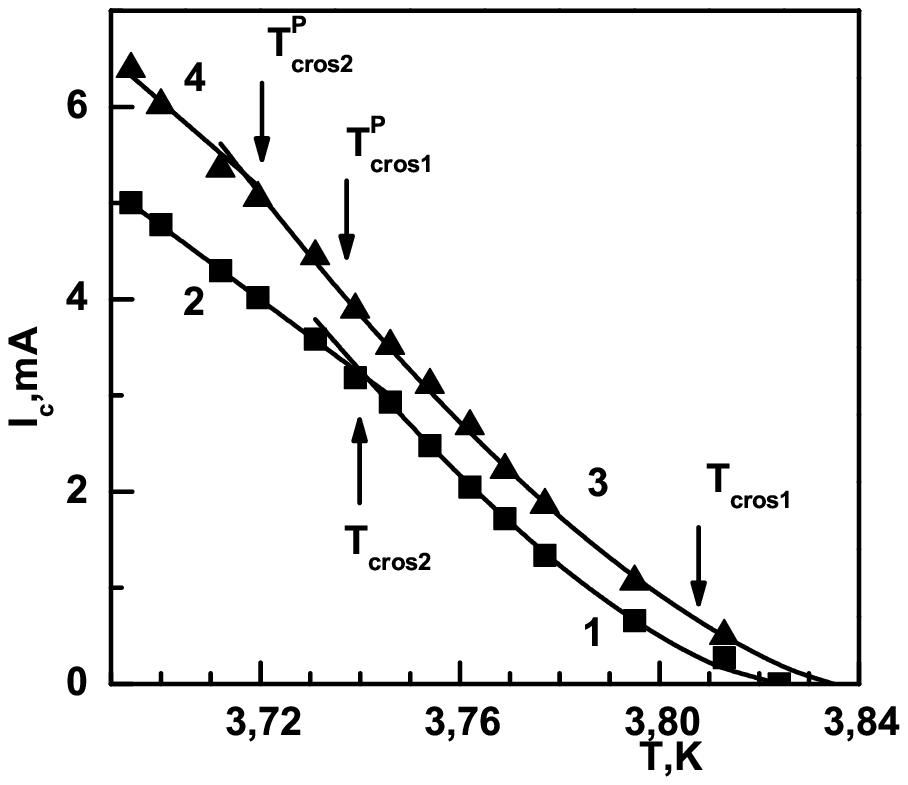}}
\caption{
%Fig.10.
The experimental temperature dependencies of the
critical currents $I_{c}(P=0)$~($\fullsquare$),
$I_{c}(f = 15.2~GHz)$~($\blacktriangle$) for the sample SnW8:
calculated dependence
$I_{c}(T)=1.0\times 10^{3}(1 - T/T_{c})^{3/2}$~mA (curve 1);
theoretical dependence
$I_{c}^{AL}(T)=1.47\times 10^{2}(1 - T/T_{c})$~mA calculated by
\Eref{Eliashberg27}~\cite{AslamazovLempitsky}~(straight line 2);
theoretical dependence $I_{c}(f=15.2~GHz)$ calculated by
\Eref{Eliashberg16} normalized by
the curve 1, and fitting dependence
$I_{c}(T)=1.0\times 10^{3}(1 - T/3.835)^{3/2}$~ mA
(curve 3);
calculated dependence
$I_{c}(T)=1.72\times 10^{2}(1 - T/3.835)$~mA~(straight line 4).
} \label{f10}
\end{figure}

\Fref{f10} presents the temperature dependencies
of the critical current $I_{c}$ for the sample
SnW8~  \cite{DmitrievZolochevskiiBezuglyi1}.
First, consider behavior of $I_{c}(T)$ without an
external electromagnetic field. A width of the film
is large enough ($w$=25 $\mu$m), so this sample is
a narrow channel only in the immediate
vicinity~of $T_{c}=3.816$~K, and at $T < T_{cros1}=3.808$~K
behaves as a wide film. At $T_{cros2}=3.740~K<T<T_{cros1}$
the temperature dependence of the critical
current has the form $(1 - T/T_{c})^{3/2}$, although the
value of $I_{c}$ is less than $I_{c}^{GL}$.
At $T < T_{cros2}$ the temperature dependence of the
critical current becomes linear and corresponds to
the Aslamazov -- Lempitsky theory \cite{AslamazovLempitsky}
$I_{c}(T) = I_{c}^{AL}(T) = 1.47\times 10^{2}(1 - T/T_{c})$~mA.

In a microwave field with a frequency $f$ = 15.2~GHz, similar
to the narrow sample SnW10, there is an
increase of the critical temperature $T_{c}^{P}$(15.2~GHz)=3.835~K
and a noticeable decrease of
crossover temperatures: $T_{cros1}^{P}$= 3.738~K and
$T_{cros2}^{P}$=3.720~K. At the same time, in order
to achieve a good agreement between the experimental
dependence of the highest enhanced critical
current $I_{c}^{P}(T)$ and \Eref{Eliashberg16}, it is
necessary to normalize this formula by the
measured equilibrium ($P$=0) critical current
$I_{c}(T) = 1.0\times 10^{3}(1 - T/T_{c})^{3/2}$~mA over
the entire temperature range measured (\fref{f10}, curve 3).
In this temperature range, the
critical current can be approximated by the dependence:
$I_{c}(T) = 1.0\times 10^{3}(1 - T/3.835)^{3/2}$~mA.
When $T<T_{cros2}^{P}$ the temperature dependence
$I_{c}^{P}(T)$ is linear (\fref{f10}, straight line 4).
From the data in \fref{f10} it follows that
under enhancement of superconductivity by a microwave
field even a fairly wide film behaves as a
narrow channel down to low temperatures than that
without irradiation ($T_{cros1}^{P}<T_{cros1}$);
the vortex region in the IVC in this temperature
range is also absent \cite{DmitrievZolochevskiiBezuglyi1}.

A qualitative similarity between obtained in reference
\cite{DmitrievZolochevskiiBezuglyi1} results of studies of
wide films and experimental results of studies of narrow
channels \cite{DmitrievKhristenko}, and the ability
to quantitatively describe the temperature dependence
of $I_{c}^{P}(T)$ in wide films using equations
of the Eliashberg theory suggest that the mechanism of
the enhancement effect is common for both wide
films and narrow channels \cite{DmitrievZolochevskiiBezuglyi1}.
It lies in increasing the energy gap, caused by the redistribution
of nonequilibrium quasiparticles to higher
energies under a microwave field \cite{Eliashberg1}.
This conclusion is not entirely obvious for wide
films with a non-uniform current distribution
across the sample width.

A similarity of enhancement mechanisms in narrow
channels and wide films can be supported by the
following observations. Despite an increase of the
current density, near the edges of a wide film the
main current, both transport and induced by a microwave
field, is distributed over the entire film
width. Thus, the nonequilibrium of quasiparticles
in a wide film, as in a narrow channel, is excited
by a microwave field within an entire volume of a
superconductor, and therefore, the effect of
enhancement in wide films undergoes a certain
quantitative modification due to non-uniform
current distribution. In this regard, we emphasize a
significant difference in the conditions of
formation of nonequilibrium under a microwave field in
wide films and bulk superconductors, in
which up to now the effect of enhancement was not observed.
In the latter case, the total current
is concentrated in a thin Meissner layer of the
thickness of $\lambda$ near the surface of the metal,
which leads to an additional relaxation mechanism:
the spatial diffusion of nonequilibrium
quasiparticles excited by the microwave field, from the
surface into the equilibrium volume. An
intensity of this mechanism is determined by time
$\tau_{D}=\lambda^{2}(T)/D$ during which quasi-particles
leave Meissner layer, which at typical temperature is three to four
orders of magnitude less than the inelastic relaxation time.
This high efficiency of the diffusion
mechanism of relaxation, apparently, leads to suppression
of the effect of enhancement in bulk
superconducting samples \cite{DmitrievZolochevskiiBezuglyi1}.

The current state of a wide film, despite the
semblance with the Meissner state of a bulk
current-carrying superconductor is qualitatively
different from the latter. While in a bulk
superconductor the transport current is concentrated
in a thin surface layer and decays
exponentially at a distance of the London penetration
depth $\lambda(T)$  away from the surface,
in a wide film the current is distributed over its
entire width $w$ according to approximating power
law $[x(w - x)]^{-1/2}$, where $x$ is the transverse
coordinate \cite{AslamazovLempitsky}. Thus,
the characteristic length $\lambda_{\bot}(T) = 2\lambda^{2}(T)/d$
($d$ is the film thickness), called
in the theory of the current state of wide films
usually as a penetration depth of perpendicular
magnetic field, in fact defines not a spatial scale of
the current decay for outgoing from the edges,
but a magnitude of the edge current density, acting
as a "cut-off" factor in the
above-mentioned law of the current density distribution
at distances $x, w - x \sim \lambda_{\bot}$
away from the film edges \cite{DmitrievZolochevskiiBezuglyi1}.

Being based on a qualitative difference between the
current states in bulk and thin-film superconductors,
one can argue that moderate non-uniformity of the
current distribution in wide films does not cause
fatal consequences for the effect of enhancement,
and that the diffusion of nonequilibrium
quasiparticles excited in the whole bulk of a film,
making only minor quantitative deviations from
the Eliashberg theory \cite{DmitrievZolochevskiiBezuglyi1}.

The authors of reference \cite{DmitrievZolochevskiiBezuglyi1}
used a modeling approach to account for these deviations
by introducing a numerical form factor of the current
distribution in \Eref{Eliashberg16} for the enhanced critical
current of Eliashberg. They evaluated this
form factor by fitting the limiting case of
\Eref{Eliashberg16} at zero-power microwave
irradiation, i.e., \Eref{Eliashberg17} for measured
values of the equilibrium critical current.
They then used the obtained values of the form factor
in \Eref{Eliashberg16} for $P\neq0$, to yield
a good agreement with the experimental data
(see \fref{f9}).

Noteworthy is the question of how the Eliashberg's
mechanism "works" in a wide film in narrow channels
the superconductivity is destroyed due to the
mechanism of homogeneous pair-breaking of Ginzburg-Landau,
while in a wide film the superconductivity is destroyed
due to the emergence of vortices. We believe that
in this case the enhancement of the energy gap
leads to a corresponding increase of the barrier for
entering vortices, and that enhances the critical
current in a wide film
\cite{DmitrievZolochevskiiBezuglyi1}. It is
interesting to note that there are no significant features
in the curves $I_{c}(T)$, when upon decreasing the
temperature the films go from the regime of a
narrow channel to the regime of a wide vortex
film. One can therefore conclude that the transition
between the regimes of a uniform pair-breaking and
vortex resistivity affect neither the value nor
the temperature dependence of the critical current.

To complete the discussion of the effect of
superconductivity enhancement, we attract an attention
to the empirical fact that all the theoretical
curves for $I_{c}^{P}(T)$, derived from the equations
of the Eliashberg theory, are well approximated
by a power law $(1-T/T_{c}^{P})^{3/2}$. This law is
very similar to the temperature dependence of
the pair-breaking current of Ginzburg-Landau, in which
the critical temperature $T_{c}$ is replaced with its
enhanced value $T_{c}^{P}$. The explicit expressions
for such approximating dependencies with numerical
coefficients are given in the captions to \fref{f9}
and \fref{f10}.

The other important result of these studies is
significant expansion of temperature range near
the superconducting transition temperature,
where a film behaves as a narrow channel during
enhancement of superconductivity: In a microwave
field, the crossover temperature in the wide film
regime $T_{cros1}^{P}$ is significantly reduced
compared to its equilibrium value $T_{cros1}$,
while $T_{c}^{P}>T_{c}$. At first glance, this
result is somewhat contrary to the criterion of the
transition between the different regimes of
a superconducting film:
$w = 4\lambda_{\bot}(T_{cros1})$
\cite{DmitrievZolochevskiiBezuglyi} since
an increase of the energy gap upon irradiation
implies a reduction of the $\lambda_{\bot}$ and,
consequently, reduction
of the characteristic size of vortices. This obviously
makes easier the conditions for entry of vortices
into a film. Consequently, the crossover temperature
is expected to increase in a microwave field.
However, it appears that the mechanism of an
influence of microwave radiation on vortices is
somewhat different. So, it was found that a wide
film with a vortex region in IVC under a microwave
field behaves like a narrow channel: The vortex
region in IVC disappears (see, e.g., \fref{f9},~
($\blacktriangledown$)). It should be noted that
this kind of IVC may be in two cases. In the first
case, under the influence of a microwave field there
is delay in motion of vortices up to the point
of their termination, i.e., vortices appear, but
under a microwave radiation, they do not move.
In the second case the vortices do not appear at
all. Turn back to \fref{f9}. In the temperature range
$T_{cros1} = 3.769~K <T < T_{c} = 3.809~K$ without
microwave irradiation the sample SnW10 is a narrow
channel. Under microwave field with the frequency
$f$=12.9~GHz in this sample there is an increase
of the critical current (enhancement of superconductivity)
(see \fref{f9}~($\blacktriangledown$)).
At the same time, it is important to emphasize
that the temperature dependence of the enhanced
critical current agrees well with the theoretical
dependence $I_{c}^{P}(T)$ (see \fref{f9},
curve 5), plotted in accordance with the Eliashberg
theory for a narrow superconducting channel with
a uniform current distribution over the cross section
of the sample. It is interesting to note that
this theoretical dependence coincides well with
experimental points (see \fref{f9}
($\blacktriangledown$)) not only in the
temperature range $T_{cros1}=3.769~K<T<T_{c}= 3.809$~K,
in which the sample SnW10 is a narrow channel in the absence
of microwave irradiation, but at much lower
temperatures (up to $T<3.700$~K). This behavior of
$I_{c}^{P}(T)$ suggests that under the microwave field
of $f$ = 12.9~GHz in the temperature range
$3.700~K<T<T_{c}^{P}$ there are no vortices in the sample
SnW10. Otherwise, in \fref{f9}, in the temperature range
$3.700~K < T < 3.769~K$ the values of
$I_{c}^{P}(T)$ would be lower compared with the
theoretical curve 5 calculated within the Eliashberg
theory. Moreover, the crossover would be seen in
the dependence $I_{c}^{P}(T)$ upon entering vortices.
The suppression of vortex resistivity in a wide film
by a microwave field is discussed in more detail
in reference \cite{DmitrievZolochevskiiSalenkovai25}.

Thus, referring to \fref{f9}, one can say the following.
For the irradiation frequency $f$=12.9~GHz
the maximum value of $I_{c}^{P}(T)$ in the
sample SnW10 is realized at high ($P/P_{c}$=0.45) power of an
external microwave field, which prevents the formation of
vortices, so the sample behaves as a narrow channel
in the temperature range from $T_{c}^{P}$ up to $T < 3.700$~K.
In this case, the curve 5 in \fref{f9}, plotted
using \Eref{Eliashberg16} of the Eliashberg theory
for a narrow channel, and giving the pair-breaking
current density of Ginzburg -- Landau at $P$=0 is in
a good agreement with the experimental curve
$I_{c}^{P}(T)$  (\fref{f9},~($\blacktriangledown$)).
On this basis, it can be argued that in this case due
to a microwave field the sample becomes a narrow channel
(there is no vortex region in IVC and $I_{c}^{P}(T)$
is fully consistent with the formula \eref{Eliashberg16}
of the Eliashberg theory, assuming a uniform
distribution of the superconducting current
over the cross section of the sample.

As the irradiation frequency ($f$=9.2~GHz) decreases
the power at which the maximum value of $I_{c}^{P}(T)$
is realized and consequently its influence are reduced.
This leads to a smaller decrease of $I_{cros1}^{P}$
with respect to $T_{cros1}$. It is important to note
that in this case too, the experimental dependence
$I_{c}^{P}(T)$ (see \fref{f9}, (\fullcircle))
agrees quite well with the curve 4 in \fref{f9},
plotted according to \Eref{Eliashberg16} for a narrow
channel, up to the temperature of
$T^{**} = 3.760~K < T_{cros1} = 3.769~K$. At temperatures
$T_{cros1}^{P} < T < T^{**}$ for the sample
SnW10 there is no vortex region in IVC, but $I_{c}^{P}(T)$
deviates downward from the theoretical curve 4 in
\fref{f9} plotted for a narrow channel and normalized in
such a way that it gives the pair-breaking
current of Ginzburg -- Landau at $P$=0.

As can be seen from \Eref{Eliashberg17} and \Eref{Eliashberg18}
in the Eliashberg theory the expression for
the critical current at $P$=0, enhanced by a microwave
field, transforms to the formula for the
pair-breaking current of Ginzburg -- Landau. Like the whole
theory, this is true only in the case of a
narrow channel. At the same time, at temperatures
$T < T_{cros1}$ the SnW10 film reveals itself as wide
(there appear a vortex region in IVC), the distribution
of the superconducting current over its cross
section becomes non-uniform, and the critical current
$I_{c}(T) = 5.9\times 10^{2}(1 - T/T_{c})^{3/2}$~mA of
this film at $P$ = 0 is less than the pair-breaking current
$I_{c}^{GL}(T)  = 7.07\times 10^{2}(1 - T/T_{c})^{3/2}$~mA
for the sample SnW10, although the
temperature dependence is preserved. Interestingly
that the ratio $I_{c}(T)/I_{c}^{GL}(T)\simeq0.83$ for
sample SnW10. In this case, it turns
out that if a normalization factor in
\Eref{Eliashberg16} is introduced so that at $P$=0 it will
give not $I_{c}^{GL}(T)$ but $I_{c}(T)$, then
using the formula one can plot a curve (see \fref{f9}, curve 6),
which is in a good agreement with the
experimental dependence $I_{c}^{P}(T)$. It should be noted
that in this case, the normalization factor is
0.83 \cite{DmitrievZolochevskiiBezuglyi1}. An introduction
of a universal normalization factor over the
entire temperature range $T_{cros2} < T < T_{cros1}$ is
possible due to the fact that the temperature
dependence $I_{c}^{P}(T)$, described by \Eref{Eliashberg16},
although is quite complex, yet is numerically
very close to the law $\propto (1-T/T_{c}^{P})^{3/2}$, which
at $P$=0 transforms to the dependence
$I_{c}(T)\varpropto ( 1-T/T_{c})^{3/2}$ for a wide film.

A similar situation is observed for a much wider film SnW8.
This sample is a narrow channel only in the
immediate vicinity of the $T_{c}$. Therefore, for temperatures
$T < T_{cros1}=3.808$~K \Eref{Eliashberg16}
gives the values of the enhanced critical current that do not
coincide with the experimental values of
$I_{c}^{P}(T)$. However, for normalization of \Eref{Eliashberg16} to
the equilibrium critical current
$I_{c}(T) = 1.0\times 10^{3}(1 - T/T_{c})^{3/2}$~mA
at $P$=0 there is also a good agreement between theory
and experiment (\fref{f10}, curve 3). Note that in this case
too, the normalization factor of
\Eref{Eliashberg16} of the Eliashberg theory is the same as
the ratio $I_{c}(T)/I_{c}^{GL}(T)$.

Thus, we conclude that if the equilibrium critical current
($P$=0) of a wide film has the temperature
dependence $I_{c}(T)\varpropto (1 - T/T_{c})^{3/2}$,
typical for a narrow channel, then using the formula of
the Eliashberg theory, normalized to $I_{c}(T)$, one can
well describe the experimentally measured dependencies
of the enhanced critical current $I_{c}^{P}(T)$, which
are numerically very close to $(1-T/T_{c}^{P})^{3/2}$.
In the temperature range $T < T_{cros2}^{P}$, where the
temperature dependence of the critical current for a
wide film is linear, $I_{c}(T)\varpropto 1 - T/T_{c}$, the
temperature dependence of the enhanced critical
current is also linear: $I_{c}^{P}(T)\varpropto 1 - T/T_{c}^{P}$.
These facts, albeit indirectly, confirm
the hypothesis that the mechanism of superconductivity enhancement
in wide films is the same as in narrow channels.
\subsection{\label{sec:CURPSC}The current of phase-slip processes}
This section presents the results of experimental study
of enhancement of the current $I_{m}(T)$, at which
the first PSL is formed, in a wide temperature range
under an external microwave irradiation of
different frequencies \cite{DmitrievZolochevskii23}.

Taking into account the fact that there is no theory
of superconductivity enhancement in wide films, one can
try at least qualitatively to describe the effect of
microwave irradiation on the current $I_{m}(T)$ using
the following considerations. In studies of the critical
current enhancement in superconducting films
with different width, the following experimental facts
were obtained \cite{DmitrievZolochevskiiBezuglyi1}.
In narrow channel, the equilibrium critical current
has the temperature dependence
$I_{c}^{GL}(T)\varpropto ( 1 - T/T_{c})^{3/2}$. At
the same time, the enhanced critical current
$I_{c}^{P}(T)$  of this channel, perfectly
described by the Eliashberg theory
\cite{Eliashberg1,IvlevEliashberg,Eliashberg2,IvlevLisitsyn,Schmid}
can be well approximated by the dependence
$I_{c}^{P}(T)\varpropto (1-T/T_{c}^{P})^{3/2}$
\cite{KlapwijkBerghMooij}. Here, $T_{c}^{P}$
is the enhanced critical temperature. In a wide (vortex)
film near $T_{c}$ the temperature dependence of
the equilibrium critical current is
$I_{c}(T)\varpropto (1 - T/T_{c})^{3/2}$
\cite{DmitrievZolochevskii}. It turns out that the critical
current enhanced by a microwave field
in this case can also be well approximated by a similar dependence:
$I_{c}^{P}(T)\varpropto (1-T/T_{c}^{P})^{3/2}$
\cite{DmitrievZolochevskiiBezuglyi1}. When
$T < T_{cros2}$ in a wide film there is a linear temperature
dependence of the equilibrium critical
current \cite{DmitrievZolochevskii}. Almost at the same
temperatures the enhanced critical
current can also be approximated by the linear dependence
$I_{c}^{P}(T)\varpropto (1-T/T_{c}^{P})$
\cite{DmitrievZolochevskiiBezuglyi1}. Based on the
above experimental facts, one can also try to approximate
the temperature dependencies of the current $I_{m}^{P}(T)$,
enhanced by a microwave field, by a dependence
similar to \Eref{Eliashberg22} for the equilibrium case.

\begin{figure}
\centerline{\includegraphics[height=3in]{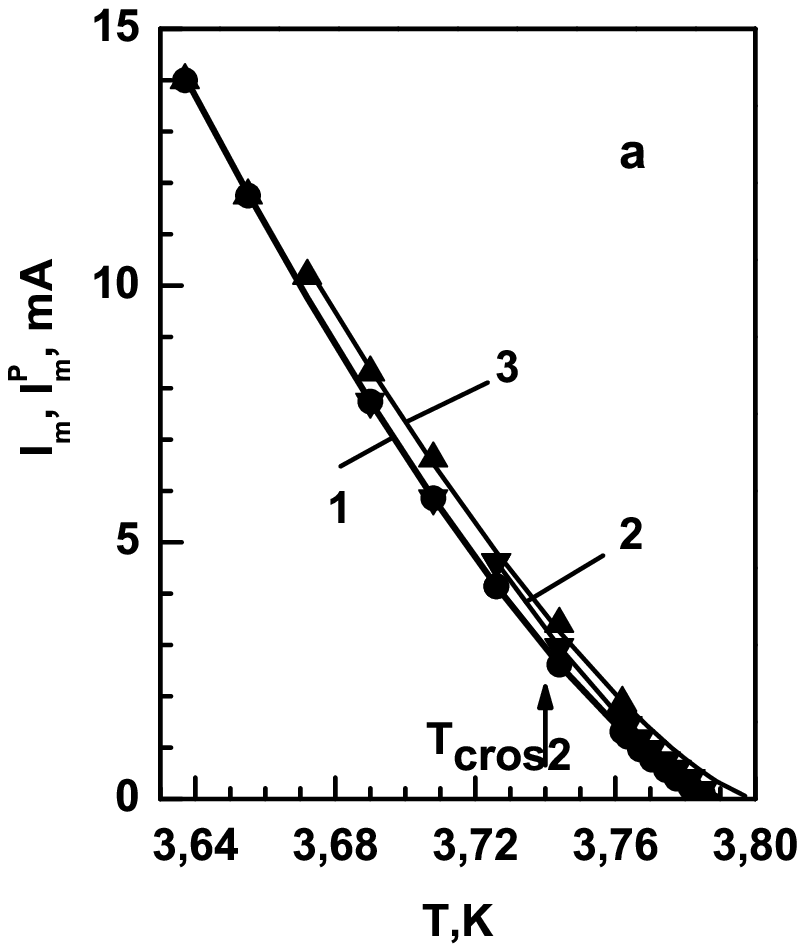}}
\centerline{\includegraphics[height=3in]{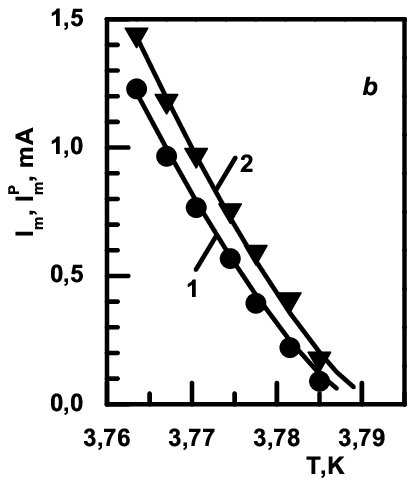}}
\caption{ \label{f11}
%Fig.11.
The experimental temperature dependencies of the
maximum current $I_{m}$ of existence of stationary
uniform flow of intrinsic vortices of transport
current across the film SnW5:
$I_{m}(T,P = 0)$ (\fullcircle),
$I_{m}^{P}(T,f=9.2~GHz)$~($\blacktriangledown$) and
$I_{m}^{P}(T,f=15.2~GHz$)~($\blacktriangle$).
Curve 1 is the theoretical dependence $I_{m}(T)$
(see \Eref{Eliashberg23}); curve 2 is the calculated
dependence $I_{m}^{P}(T,f=9.2~GHz$) (see \Eref{Eliashberg24});
curve 3 is the calculated dependence
$I_{m}^{P}(T,f=15.2~GHz$) (see \Eref{Eliashberg26}).}
\end{figure}

\Fref{f11} shows the experimental temperature dependencies
of the currents $I_{m}^{P}(T)$ in a microwave
field and currents $I_{m}(T)$ in the absence of
the field for the sample SnW5 \cite{DmitrievZolochevskii23}.
For clarity, in \fref{f11}(b) the results are shown for a
narrower temperature range near $T_{c}$ than that in
\fref{f11}(a). A width of the film SnW5 is large enough
($w$~=~42~$\mu$m), so even for temperatures
$T < T_{cros2}=3.740$~K there is a linear temperature
dependence of the critical current
\cite{DmitrievZolochevskii}, what is close enough to $T_{c}$.

First, we consider the behavior of the current at which the
first PSL appears, $I_{m}(T)$, without an
external electromagnetic field (see \fref{f11}, (\fullcircle)).
The solid curves 1 in these figures are
calculations of $I_{m}(T)$ according to \Eref{Eliashberg22}
with taking into account the film
parameters (see Table 1)
\begin{eqnarray}\label{Eliashberg23}
I_{m}(T)= 2.867\times 10^{3}(1 - T/T_{c})^{3/2}
\times 1.35 [ \ln (2 \times 42   \\  \nonumber
\times (1 - T/T_{c})/0.02532)]^{-1/2}~[mA].
\end{eqnarray}
As can be seen in \fref{f11}, the experimental
dependence $I_{m}(T)$ is in a good agreement with
calculated one (see curve 1 )
\cite{DmitrievZolochevskii23}. The experimental
dependence of the current $I_{m}^{P}(T)$ at the
irradiation frequency $f$= 9.2~GHz (see \fref{f11},
($\blacktriangledown$)) is well approximated by the dependence
\begin{eqnarray}\label{Eliashberg24}
I_{m}^{P}(T)= 2.869\times 10^{3}(1 - T/T_{c1}^{P})^{3/2}
\times 1.44 [ \ln (2 \times 42   \\  \nonumber
\times (1 - T/T_{c1}^{P})/0.02531)]^{-1/2}~[mA].
\end{eqnarray}
similar to \Eref{Eliashberg22} (\fref{f11}, curve 2).
Here and in the calculation of the
pair-breaking current of Ginzburg -- Landau,
the enhanced critical temperature
$T_{c1}^{P}$= 3.791~K was used. The experimental
dependence of the current $I_{m}^{P}(T)$
at the frequency of an external electromagnetic
field $f$ = 12.9~GHz (not shown due to space
limitations) is well approximated by
\begin{eqnarray}\label{Eliashberg25}
I_{m}^{P}(T)= 2.875\times 10^{3}(1 - T/T_{c3}^{P})^{3/2}
\times 1.28 [ \ln (2 \times 42   \\  \nonumber
\times (1 - T/T_{c3}^{P})/0.02529)]^{-1/2}~[mA].
\end{eqnarray}
Here, the enhanced critical temperature $T_{c3}^{P}$=3.797~K
was also used. The experimental dependence
of the current $I_{m}^{P}(T)$ at the frequency of
microwave field $f$=15.2~GHz (see \fref{f11},
($\blacktriangle$)) is well approximated by
\begin{eqnarray}\label{Eliashberg26}
I_{m}^{P}(T)= 2.877\times 10^{3}(1 - T/T_{c2}^{P})^{3/2}
\times 1.28 [ \ln (2 \times 42 \\  \nonumber
\times (1 - T/T_{c2}^{P})/0.02528)]^{-1/2}~[mA].
\end{eqnarray}
similar to \Eref{Eliashberg22} (\fref{f11}(a), curve 3).
Here, the stimulated critical temperature
$T_{c2}^{P}$ = 3.799~K was used.
For measurements of the current $I_{m}^{P}(T)$ of
films in microwave field the irradiation power was chosen
in such a way that the critical current $I_{c}^{P}(T)$
was maximal. In this case the current $I_{m}^{P}(T)$
was also maximal due to some correlation of these
quantities \cite{DmitrievZolochevskiiSalenkova2}.

Since the theory \cite{AslamazovLempitsky}, in which
the definition $I_{m}(T)$ is introduced, assumes a
linear temperature dependence of the critical current
\begin{eqnarray}\label{Eliashberg27}
I_{c}^{AL}(T)= 1.5I_{c}^{GL}(0)(\pi
\lambda_{\bot}(0)/w)^{1/2}(1 - T/T_{c}).
\end{eqnarray}
then, strictly speaking, \Eref{Eliashberg22} should
be applicable only in the temperature range
$T < T_{cros2}$, where such a dependence of the
critical current is observed. However, as seen in
\fref{f11}, \Eref{Eliashberg22} for the
equilibrium dependence $I_{m}(T)$ and
\Eref{Eliashberg24}--\Eref{Eliashberg26}
for the case of enhancement of $I_{m}^{P}(T)$ by an
electromagnetic field sufficiently well
describe the experimental dependencies in the case of
$T > T_{cros2}$ too. This is obviously due to
the fact that at $T < T_{cros2}$ and $T > T_{cros2}$
the resistive current states at
$I\simeq I_{m}$ differ little from each other:
both of these states are characterized by fairly
uniform current distribution over the width of
the sample due to the quite
dense filling of the film by a lattice
of vortices  \cite{DmitrievZolochevskii23}.

Thus, the experimental temperature dependencies
of the enhanced current $I_{m}^{P}(T)$ are well approximated
by \Eref{Eliashberg24}--\Eref{Eliashberg26}, similar to
formula \eref{Eliashberg22} for the equilibrium
case of the Aslamazov -- Lempitsky theory
\cite{AslamazovLempitsky}, where the critical temperature
$T_{c}$ is replaced by the enhanced critical temperature
$T_{c}^{P}$  \cite{DmitrievZolochevskii23}.

Consider the behavior of $I_{m}^{P}(T)$ of the sample
SnW5 in a microwave field with the frequency
$f$=9.2~GHz (\fref{f11} ($\blacktriangledown$)).
It can be seen that upon irradiation of the film by
microwave power there is the enhancement of
$I_{m}^{P}(T, f=9.2~GHz)$ up to $T$=3.708~K.
At temperatures $T<3.708$~K the enhancement
of $I_{m}^{P}(T)$ was not observed. For the irradiation of
the sample SnW5 by microwave field with the frequency
$f$=12.9~GHz the enhancement of
$I_{m}^{P}(T,f=12.9~GHz)$ is observed up to
$T$=3.690~K. At lower temperatures the enhancement of
$I_{m}^{P}(T)$ was not found. For the irradiation
of the sample SnW5 by microwave field with the frequency
$f$=15.2~GHz (\fref{f11}, ($\blacktriangle$))
enhancement of $I_{m}^{P}(T, f = 15.2~GHz)$ is observed
up to $T$ = 3.655~K. At temperatures $T<3.655$~K the
enhancement of $I_{m}^{P}(T)$ was not observed.
It should be noted that in \fref{f11} it is seen that
$I_{m}^{P}(T,f=15.2~GHz) > I_{m}^{P}(T,f=12.9~GHz) >
I_{m}^{P}(T,f=9.2~GHz)$.

Thus, an absolute value of $I_{m}^{P}(T)$ increases with
the irradiation frequency, and the temperature
region of enhancement of $I_{m}^{P}(T)$
is extended toward lower temperatures
\cite{DmitrievZolochevskii23}.
By the way, in the same way behaves a
critical current $I_{c}^{P}(T)$
\cite{DmitrievZolochevskiiBezuglyi1}.
Let us try to find an explanation for this.

We take into account two factors. First of all, as already noted,
in wide films at $I\simeq I_{m}$ the
current distribution across the width of the film is
close to uniform. In this case it is wise to make use
of the knowledge accumulated for narrow channels.
Second, the source of enhanced critical parameters
$I_{m}^{P}(T)$ and $I_{c}^{P}(T)$ of a superconductor
is a non-equilibrium distribution function of
quasiparticles over energy. In this case, first of
all the value of the energy gap increases
\cite{Eliashberg1,IvlevEliashberg,Eliashberg2,IvlevLisitsyn,Schmid}.

\begin{figure}
\centerline{\includegraphics[height=3in]{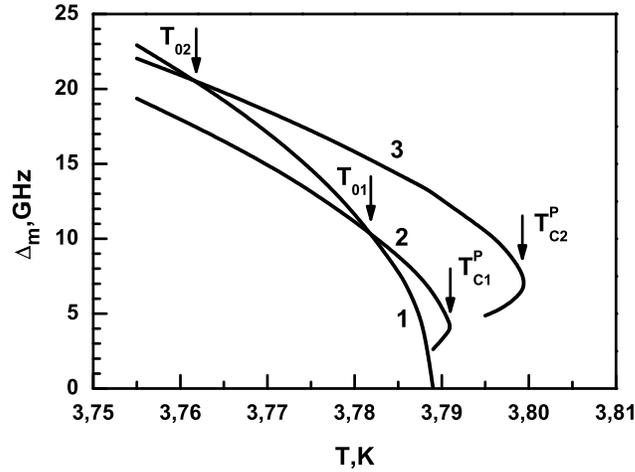}}
\caption{
%Fig.12.
The calculated dependencies of the equilibrium (curve 1) and
enhanced by microwave field the gap for the
sample SnW5 (curve 2, $f$=9.2~GHz; curve 3, $f$=15.2~GHz).
} \label{f12}
\end{figure}
With taking into account the above arguments, in
\fref{f12} the curve 1 represents a temperature dependence
of the equilibrium gap $\Delta_{0}(T)$ (in frequency units),
and the curves 2 and 3 show temperature
dependencies of the enhanced gap $\Delta_{m}^{P}(T)$
for the sample SnW5 at the irradiation frequencies of
9.2 and 15.2~GHz, assuming a uniform distribution
of the transport current density over the cross section.
In the figure it is seen that the upper branch of the
temperature dependence of the energy gap enhanced
in a superconductor, $\Delta_{m}^{P}(T)$, intersects
with a similar dependence of the equilibrium gap
$\Delta_{0}(T)$. Moreover, the higher the irradiation frequency,
the lower in the temperature is the point
of intersection of dependencies $\Delta_{m}^{P}(T)$ and
$\Delta_{0}(T)$  ($T_{02} = 3.762~K < T_{01} = 3.782~K$),
where the enhancement of the energy gap ceases.

It should be noted that the temperature ranges
where the non-equilibrium values of the energy gap and the
currents exceed the equilibrium values do not coincide.
The reason for this lies in a significant
difference between the curves of pair-breaking
$I_{s}(\Delta)$ in equilibrium and non-equilibrium
cases \cite{DmitrievKhristenko}.

It is important to note that the maximum enhanced
critical temperatures $T_{c1}^{P}= 3.791$~K at the
irradiation frequency of 9.2~GHz and $T_{c2}^{P}=3.799$~K
at the frequency of microwave field 15.2~GHz,
derived from theoretical curves 2 and 3 in \fref{f12},
are in a good agreement with the values
of $T_{c}^{P}$, obtained by fitting the
experimental curves $I_{m}^{P}(T)$ (see \fref{f11}).

\section{\label{sec:CONCL}Conclusion}

In the present review, the behavior of the critical current $I_{c}$
and the maximum current $I_{m}$ at which in
a wide film a vortex structure of the resistive state
disappears and the first phase-slip line arises
is analyzed in thin superconducting films of different
width, located in a microwave field. A enhancement
of superconductivity by an external
electromagnetic field was found in wide
$w \gg \xi(T),\lambda_{\perp}(T)$, superconducting films
\cite{AgafonovDmitriev} with a non-uniform
spatial distribution of the current over the sample width.
The superconductivity enhancement in a wide
film increases not only the critical current $I_{c}$,
but also the maximum current at which there is a
vortex resistive state, $I_{m}$ \cite{AgafonovDmitriev}. In the
framework of the Eliashberg theory an
equation for the enhanced critical current was derived and
expressed in terms of experimentally
measured quantities \cite{DmitrievZolochevskiiBezuglyi1}.
A comparison of experimental temperature dependencies of
the enhanced critical current with those calculated in
the framework of the Eliashberg theory revealed a
good agreement between them \cite{DmitrievZolochevskiiBezuglyi1}.
It was shown \cite{DmitrievZolochevskiiBezuglyi1}, that near
the superconducting transition the temperature
dependence of the enhanced critical current in not
very wide films ($w < 10\lambda_{\bot}(T)$)
appears to be numerically very close to the
law $(1 - T/T_{c}^{P})^{3/2}$  for the equilibrium
pair-breaking critical current when $T_{c}$ is replaced
with the enhanced critical temperature
$T_{c}^{P}$. It was found \cite{DmitrievZolochevskiiBezuglyi1},
that for sufficiently wide films
($w > 10\lambda_{\bot}(T)$) the enhanced critical current
has a linear temperature dependence
$I_{c}^{P}(T)\varpropto(1 - T/T_{c}^{P})$, similar to
that in the equilibrium theory of
Aslamazov -- Lempitsky with replacement of
$T_{c}$ by $T_{c}^{P}$. Experimental dependencies of the
enhanced critical current $I_{c}^{P}$ and the enhanced
current of formation of the first PSL, $I_{m}^{P}$,
on power and frequency of microwave irradiation were obtained
in thin (thickness
$d\ll\xi(T),\lambda_{\perp}(T)$) superconducting films of different width
$w$  \cite{DmitrievZolochevskiiSalenkova2, DmitrievZolochevskii23,
DmitrievZolochevskiiBezuglyi1}.
It was found experimentally that when the
film width increases, the range of irradiation power,
at which the effect of superconductivity enhancement is observed,
shrinks abruptly, and hence the probability
of its detection decreases \cite{DmitrievZolochevskiiSalenkova2}.
This statement is an answer to the question
on much delayed discovery of the enhancement effect in wide films.
It is established that with an increase of
the film width the ratio $P/P_{c}$, at which there is a maximum
enhancement effect, is reduced, and the effect
of enhancement of the critical current in wider films is observed
at lower irradiation power, since in
the microwave range the value of $P_{c}$ is practically
independent of frequency
\cite{PalsRamekers, BezuglyiDmitrievSvetlov}. The power,
at which the maximum effect of enhancement is
observed, increases with the frequency of microwave irradiation.
It was found that when the film width
increases, the curvature sign of a descending section
in the dependence $I_{c}^{P}(P)$ is
changed \cite{DmitrievZolochevskiiSalenkova2}.

Studies of the critical current enhanced by a microwave
irradiation confirmed an earlier conclusion, based
on studies of the equilibrium critical current (in the
absence of external irradiation)
\cite{DmitrievZolochevskii}, that narrow channels are
films which satisfy to the relation
$w/\lambda_{\bot}\leq4$. For them, according to a
theory \cite{Eliashberg1} the calculated values of
lower boundary frequencies of superconductivity enhancement
correspond to experimental values. In much wider
films there appears a dependence of characteristic
parameters of enhancement effect on the film width. In
this connection, to describe a non-equilibrium state of wide
films ($w/\lambda_{\bot}>4$) in
electromagnetic fields it is necessary to develop a
theory, which in contrast to the Eliashberg
theory \cite{Eliashberg1} initially takes into account a
non-uniform current distribution and the presence
of vortices of its own magnetic flux.

An unexpected effect of electromagnetic field on the
current $I_{m}$, which cannot be considered as a
trivial influence of the irradiation on the $I_{c}^{GL}(T)$ and
$\lambda_{\bot}(T)$, was
found \cite{DmitrievZolochevskii23}. The experimental temperature
dependencies of the enhanced current
$I_{m}^{P}(T)$  are well approximated by formulas which are
similar to the formula (22) for an equilibrium case
of the Aslamazov -- Lempitsky theory \cite{AslamazovLempitsky},
in which the critical temperature $T_{c}$
is replaced by the enhanced critical temperature $T_{c}^{P}$.
It was found that an absolute value of
$I_{m}^{P}(T)$ increases with the irradiation frequency,
and the temperature region of enhancement of
$I_{m}^{P}(T)$  is extended toward lower temperatures
\cite{DmitrievZolochevskii23}.

One more important fact is worth mentioning. References \cite{AgafonovDmitriev,DmitrievZolochevskiiBezuglyi1,
DmitrievZolochevskiiSalenkova2,
DmitrievZolochevskii23,DmitrievZolochevskiiSalenkovai25}
present the main results of a study
of superconductivity enhanced by microwave irradiation
in wide films. In those works, the equilibrium
critical current in wide films reached a maximum possible value -
a value of the pair-breaking current,
and corresponded to the critical current of the
Aslamazov -- Lempitsky theory. A significant excess of this
pair-breaking critical current obtained in the
Aslamazov -- Lempitsky theory in the absence of external fields
was observed under microwave irradiation. This indicates the
existence of superconductivity enhancement in wide films and
the negligible effect of overheating, if it occurs, including
an overheating of an electronic system in a superconductor.

Thus, experimental studies of films with different width
showed that the effect of superconductivity
enhancement by microwave irradiation is common,
and occurs in both the case of uniform (narrow films)
and non-uniform (wide films) distribution of the
superconducting current over the film width.


\section*{References}
\begin{thebibliography}{99}
\bibitem{WyattDmitriev}
Wyatt A F G  Dmitriev V M  Moore W S and Sheard F W 1966
{\it Phys. Rev. Lett.} {\bf 16} 1166
\bibitem{AslamazovLarkin}
Aslamazov L G and Larkin A I 1978
{ \it Sov. Phys.--JETP} {\bf 74} 2184.
\bibitem{Eliashberg1}
Eliashberg G M 1970
{\it Pis'ma Zh. Eksp. Teor. Fiz.} {\bf 11} 186
\bibitem{RudenkoKorotash}
Rudenko E M Korotash I V  Nevirkovets I P
Boguslfvskii Yu M  Zhukov P A  and Sivakov A G 1991
{\it  Supercond. Sci. Technol.} {\bf 4} 1.
\bibitem{Dmitriev1}
Dmitriev V M  Zolochevskii I V  and Khristenko E V  1993
{\it Low Temp. Phys.} { \bf 19} 249.
\bibitem{AslamazovLempitsky}
Aslamazov L G  and  Lempitsky S V 1983
{\it Zh Eksp. Teor. Fiz.} {\bf 84} 2216.
\bibitem{DmitrievZolochevskii}
Dmitriev V M and Zolochevskii I V  2006
{\it Supercond. Sci. Technol.} {\bf19}, 342
\bibitem{AgafonovDmitriev}
Agafonov A B  Dmitriev V M  Zolochevskii I V and
Khristenko E V  2001 { \it Low Temp. Phys.} {\bf 27} 686
\bibitem{IvlevEliashberg}
Ivlev B I and Eliashberg G  M  1971
{ \it Pis'ma Zh. Eksp. Teor. Fiz.} {\bf13}  464
\bibitem{Eliashberg2}
Eliashberg G M  1971
{ \it Zh. Eksp. Teor. Fiz.} {\bf 61}  1254
\bibitem{IvlevLisitsyn}
Ivlev B I  Lisitsyn S G  and  Eliashberg G M  1973
{\it J. Low Temp. Phys.} {\bf10}  449.
\bibitem{Schmid}
Schmid A  1977  { \it Phys. Rev. Lett.}  {\bf 38} 922.
\bibitem{DmitrievZolochevskiiBezuglyi}
Dmitriev V M  Zolochevskii I V and  Bezuglyi E V 2008
{ \it  Low Temp. Phys.}{\bf 34}  982
\bibitem{KlapwijkBerghMooij}
Klapwijk T M  van der Bergh J N  and  Mooij J E  1977
{\it J. Low Temp. Phys.}{\bf 26}  385
\bibitem{DmitrievKhristenko}
Dmitriev V M and  Khristenko E V  1978
{ \it Sov. J. Low Temp. Phys.} {\bf 4}  387
\bibitem{PalsRamekers}
Pals J A and  Ramekers J J  1982
{\it  Phys. Lett. A} {\bf 87}  186
\bibitem{DmitrievGubankovNad}
Dmitriev V M  Gubankov V N  and Nad' F Ya  1986
{ \it Mod. Probl. Condens. Matter Sci}. {\bf 12} 163
\bibitem{DmitrievKhristenko2}
Dmitriev V M and  Khristenko E V  1979
{\it J. Phys. Lett. (Paris)} {\bf 40}  L85
\bibitem{DmitrievZolochevskiiBezuglyi1}
Dmitriev V M  Zolochevskii I V  and  Bezuglyi E V  2006
{\it Supercond. Sci. Technol.} {\bf 19}, 883
\bibitem{DmitrievZolochevskiiSalenkova2}
Dmitriev V M  Zolochevskii I V  Salenkova T V
and Khristenko E V 2005 { \it Low Temp. Phys.} {\bf 31} 957
\bibitem{BezuglyiDmitrievSvetlov}
Bezuglyi E V   Dmitriev V M   Svetlov V N  Churilov G E
and Azovskii A Yu 1987 { \it Sov. J. Low Temp. Phys.} {\bf 13} 517
\bibitem{DmitrievZolochevskii23}
Dmitriev V M and  Zolochevskii I V 2007
{\it Low Temp. Phys.}  {\bf 33} 647
\bibitem{DmitrievZolochevskiiSalenkovai25}
Dmitriev V M  Zolochevskii I V  and
Salenkova T V 2009 { \it Low Temp. Phys. } {\bf35} 849
\end{thebibliography}
\end{document}